\documentclass[10pt,twocolumn,journal]{IEEEtran}
\usepackage{latexsym}
\usepackage{float}
\usepackage{stfloats}
\usepackage{pifont}
\usepackage{bbding} 
\usepackage{amsmath}
\usepackage{amsfonts}
\usepackage{amsbsy}
\usepackage{amssymb}
\usepackage{times}
\usepackage{graphicx}
\usepackage{setspace}
\usepackage{enumerate}
\usepackage[usenames]{color}
\usepackage[dvips]{pstcol}
\usepackage{epstopdf}
\usepackage{caption}
\usepackage{cite}
\usepackage{amssymb}
\usepackage{amsfonts}
\usepackage{graphicx}
\usepackage{epsfig}
\usepackage{psfrag}
\usepackage{xcolor}
\usepackage{amsfonts, bm}
\usepackage{epstopdf}
\usepackage{cite}
\usepackage{color}
\usepackage{xcolor}
\usepackage{subfig}
\usepackage{verbatim}
\usepackage{multirow}
\usepackage{array}
\usepackage{booktabs}
\usepackage{amsthm}
\usepackage{makecell}
\usepackage{units}
\usepackage[linesnumbered, ruled]{algorithm2e}
\usepackage{algpseudocode}
\usepackage[ruled]{algorithm2e}
\allowdisplaybreaks[4] 

\newtheorem{assumption}{Assumption}

\newtheorem{remark}{Remark}

\newtheorem{definition}{Definition}

\IEEEoverridecommandlockouts
\begin{document}
\title{ROME: Robust Model Ensembling for Semantic Communication Against Semantic Jamming Attacks}
\author{\IEEEauthorblockN{ Kequan Zhou, Guangyi Zhang, Yunlong Cai, Qiyu Hu, and Guanding Yu}
	\thanks{ K. Zhou, G. Zhang, Y. Cai, Q. Hu, and G. Yu are with the College of Information Science and Electronic Engineering, Zhejiang University, Hangzhou 310027, China (e-mail: kqzhou@zju.edu.cn; zhangguangyi@zju.edu.cn; qiyhu@zju.edu.cn; ylcai@zju.edu.cn; yuguanding@zju.edu.cn).} }


\maketitle
\vspace{-3.3em}
\begin{abstract}
Recently, semantic communication (SC) has garnered increasing attention for its efficiency, yet it remains vulnerable to semantic jamming attacks.
These attacks entail introducing crafted perturbation signals to legitimate signals over the wireless channel, thereby misleading the receivers' semantic interpretation.
This paper investigates the above issue from a practical perspective.
Contrasting with previous studies focusing on power-fixed attacks, we extensively consider a more challenging scenario of power-variable attacks by devising an innovative attack model named Adjustable Perturbation Generator (APG), which is capable of generating semantic jamming signals of various power levels.
To combat semantic jamming attacks, we propose a novel framework called Robust Model Ensembling (ROME) for secure semantic communication.
Specifically, ROME can detect the presence of semantic jamming attacks and their power levels.
When high-power jamming attacks are detected, ROME adapts to raise its robustness at the cost of generalization ability, and thus effectively accommodating the attacks.
Furthermore, we theoretically analyze the robustness of the system, demonstrating its superiority in combating semantic jamming attacks via adaptive robustness. 
Simulation results show that the proposed ROME approach exhibits significant adaptability and delivers graceful robustness and generalization ability under power-variable semantic jamming attacks.
\end{abstract}
\IEEEpeerreviewmaketitle

\section{Introduction}\label{Introduction}
In recent times, there has been a surge in developing semantic communication (SC) systems based on deep neural networks (DNNs) \cite{Bourtsoulatze2019deep, Zhang2024a, Xie2021deep, Wang2023wireless, Zhao2024joint, Zhou2024feature}.
{The basic idea of SC is to extract the semantics from source messages and interpret them at a destination, where only the most task-relevant semantics need to be transmitted.
This process involves an end-to-end design of DNN-based systems, tailored to perform specific tasks such as classification \cite{Sagduyu2024will}.
Despite their promising performance, these systems encounter various security challenges, including information security and semantic machine learning (ML) security \cite{Yang2024secure}.
Within the realm of semantic ML security, semantic jamming attacks have emerged as a significant threat to SC systems, attributed to the inherent vulnerability of DNNs \cite{Madry2018towards}.}
Semantic jamming entails introducing crafted imperceptible perturbation signals to legitimate signals over the wireless channel, so as to mislead the DNNs at the legitimate receiver \cite{Hu2023robust}.
Once an SC system is attacked and paralyzed, the cost could be catastrophic, especially in future applications like autonomous driving.
Hence, it is imperative to thoroughly examine semantic jamming attacks in wireless communications and to explore effective countermeasures actively.
\subsection{Prior Work}
{Jamming attacks aim to interfere with legitimate wireless communications, where traditional methods typically focus on signal-level distortion \cite{Grover2014jamming}.
Specifically, energy suppression is a prevalent approach employed in elementary jamming attacks, such as constant jamming \cite{Xu2005feasibility}.
To be more energy efficient, some reactive jamming methods \cite{Pelechrinis2010denial} were designed to monitor the network and emit jamming only upon observing communication activities on specific channels.
Furthermore, some advanced jamming attacks were designed to conserve energy while magnifying their jamming effects, such as the pulsed-noise jamming \cite{Muraleedharan2006jamming}.

Although conventional jamming techniques have posed threats to wireless communications, they are primarily aimed at distorting physical signals rather than semantic content.
In contrast, semantic jamming attacks take into account the semantics of the transmitted data to mislead the legitimate receiver's interpretation, presenting a new challenge in wireless communications.
While the studies in \cite{Flowers2019communications,Restuccia2020generalized,Kim2020over}  delved into the implementation of semantic jamming attacks, they assumed that the signal to be perturbed is known to the attacker, which is impractical in real-time wireless attack scenarios.
To bridge this gap, the authors in \cite{Sadeghi2019physical} proposed a universal adversarial perturbation (UAP) attack, which does not require knowledge of the target signal.
However, this approach generates a fixed UAP vector, which can be readily estimated and subtracted from the received signal \cite{Bahramali2021robust}.
To address this issue, the authors in \cite{Bahramali2021robust} developed a perturbation generator model (PGM), capable of producing random UAP vectors that are resilient to subtraction defenses.
Additionally, the authors in \cite{Nan2023physical} proposed SemAdv, a random UAP attack model that misleads the legitimate receiver to interpret the semantics as the attacker's desired results.
Nevertheless, both PGM and SemAdv are designed on a fixed jamming power configuration, thus failing to accommodate diverse jamming power requirements during deployment.

To combat semantic jamming attacks, several countermeasures have been recently proposed \cite{Nan2023physical, Tang2023gan, Liu2023semprotector, Ren2023asymmetric}.
A prevalent strategy is adversarial training (AT) \cite{Goodfellow2014explaining}, which introduces adversarial attacks during the training stage to improve model resilience against attacks.
While this technique can enhance a model's robustness against semantic jamming, it often comes at the expense of reduced performance under non-attack conditions.
This is because DNNs are recognized to possess an intrinsic trade-off between robustness and generalization ability \cite{Tsipras2018robustness}.
Here, ``robustness'' denotes a model's ability to sustain performance under adversarial attacks, while ``generalization ability'' refers to its performance across unseen data.
To address this issue, the authors in \cite{Nan2023physical} proposed a hybrid AT technique, incorporating clean samples into the AT process to preserve both robustness and accuracy.
However, this approach achieves only average performance since the model's parameters become fixed after training and the model fails to dynamically adapt to the presence or absence of attacks.
Instead, the authors in \cite{Tang2023gan} proposed a distinct strategy, drawing inspiration from generative adversarial networks (GANs).
They integrated a jamming detector with the legitimate receiver, enabling the identification of jamming attacks.
Additionally, the work presented in \cite{Liu2023semprotector} also proposed a defense strategy based on the GAN framework, incorporating a semantic signature vector to further safeguard the transmitted semantics.
Nonetheless, both of these GAN-inspired methods only utilized the detector model for offline training, and did not leverage its detection capabilities to guide the receiver's online operations.
Thus, these approaches could only attain moderate online performance.
Furthermore, real-world attack scenarios are far more complex than the aforementioned binary case of attack presence or absence.
In fact, the power of semantic jamming signals can vary significantly, where increased jamming power allows for crafting more disruptive perturbations that could further impair the DNN's performance.
Consequently, the performance of a DNN model degrades as the attack's power level increases.
This reveals substantial weaknesses in prior defense strategies that lack online adaptive countermeasures.}

\subsection{Motivation and Contributions}
{While prior studies have made considerable progress, several limitations continue to hinder their practical application.
\begin{itemize}
	\item \textbf{\textit{Limitations in Attack Methods:}}
	Existing attack strategies are constrained by fixed jamming power settings, limiting their adaptability in real-world scenarios.
	In practice, attackers often need the flexibility to adjust their jamming power to achieve varying objectives, such as reducing power to enhance stealth or increasing it to maximize attack effectiveness.
	However, a mismatch in jamming power between the model's training and deployment phases can severely degrade the attack's performance.
	This is because power constraints are typically imposed at the end of the process, preventing the model from accounting for these constraints during the jamming generation phase.
	Consequently, the attack model's effectiveness heavily relies on maintaining consistent jamming power settings across both training and deployment phases.
	\item \textbf{\textit{Limitations in Defense Strategies:}}
	While current defense strategies exhibit certain effectiveness, their performance remains average due to inherent shortcomings.
	These approaches are largely passive, relying heavily on offline AT processes.
	Notably, in the absence of attacks, adversarially trained models are often outperformed by standard models due to the inherent trade-off in DNNs between robustness and generalization ability.
	Moreover, real-world scenarios are significantly more complex, with jamming power levels varying intricately according to the attacker's configurations.
	Consequently, existing defense strategies lacking online adaptive countermeasures struggle to accommodate the dynamic nature of real-world attacks.
	As illustrated in Fig. \ref{Tradeoff}, a model robust against high-power semantic jamming attacks, denoted by the red dotted curve, typically exhibits limited generalization ability and poor performance under low-power jamming attacks.
	Furthermore, most existing defense methods rely heavily on empirical designs and lack in-depth theoretical analysis of system robustness.
\end{itemize}
\begin{figure}[h]
	\centering
	\includegraphics[width=0.26\textwidth]{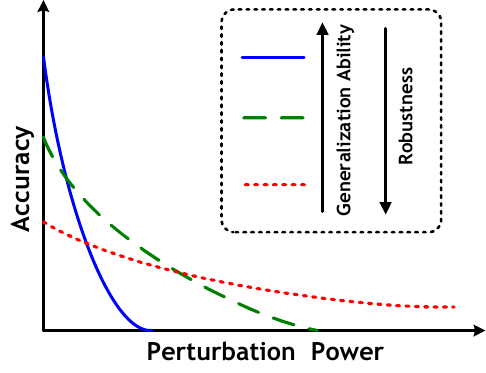}
	\captionsetup{font=footnotesize}
	\caption{The trade-off between robustness and generalization ability.}
	\label{Tradeoff}
\end{figure}
\begin{figure*}[t]
	\centering
	\includegraphics[width=0.8\textwidth]{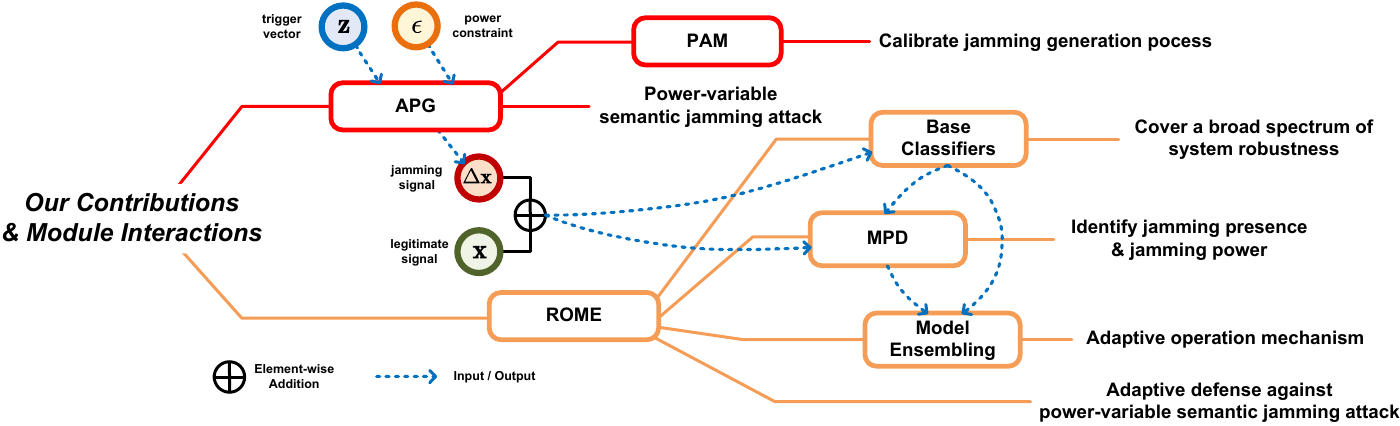}
	\captionsetup{font=footnotesize}
	\caption{Mind map of our contributions and module interactions.}
	\label{mindmap}
	\end{figure*}}

{To bridge the aforementioned gaps, this paper addresses a practical scenario characterized by power-variable semantic jamming attacks.
We concentrate on this challenge within the context of image classification tasks in semantic communications, and our approaches could be analogously applied to other data modalities and intelligent tasks.
To address the limitations of existing attack methodologies, we develop an attack model capable of accommodating different jamming power constraints, named adjustable perturbation generator (APG).
Specifically, we aim to provide the perturbation generator with the knowledge of the power constraint $\epsilon$ prior to the final power constraining step.
To this end, we devise a novel power anticipation module (PAM), which operates as a proactive calibrator.
This module allows the generator to foresee the upcoming power constraint throughout the perturbation generation process, thereby accommodating different jamming power constraints adaptively.

To address the limitations of current defense strategies, we propose a defense framework called Robust Model Ensembling (ROME) for secure semantic communication.
In light of the observation depicted in Fig. \ref{Tradeoff}, it is possible for a system to achieve better performance by dynamically balancing the trade-off between robustness and generalization ability in response to attacks of varying power levels.
To this end, we implement ROME by first ensembling a collection of base classifiers that possess distinct degrees of robustness, thereby covering a broad spectrum of system robustness for adaptive adjustment.
Then, a novel multi-level perturbation detector (MPD) is developed to simultaneously identify the presence of semantic jamming signals and measure their power levels.
With the base classifiers and MPD in place, a model ensembling algorithm is subsequently designed to adaptively ensemble the base classifiers' inference results based on the MPD's guidance.
Furthermore, we proceed to conduct a rigorous theoretical investigation to reveal the underlying mechanisms of ROME's adaptive robustness.
The key contributions of this paper are summarized as follows, with the mind map depicted in Fig. \ref{mindmap} serving as a visual aid.
\begin{itemize}
	\item We propose a novel attack model, APG, to characterize real-world semantic jamming attacks, which generates effective jamming signals covering a wide range of power levels.
	\item We develop an advanced semantic jamming detection model, MPD, which is capable of identifying the presence of semantic jamming attacks as well as measuring their power levels, allowing for adaptive countermeasures in online active defenses.
	\item We propose ROME as an innovative defense approach against semantic jamming attacks.
	This approach allows the system to adaptively balance its robustness and generalization ability, thus accommodating attacks of various power levels, alongside scenarios without attacks.
	\item  We conduct a thorough theoretical analysis to assess the robustness of ROME, highlighting its effectiveness in countering power-variable semantic jamming attacks.
	Simulation results validate our analysis, demonstrating that ROME exhibits significant adaptability and achieves graceful trade-off.
\end{itemize}}

\subsection{Organization and Notations}
The rest of the paper is organized as follows.
Section \ref{System} introduces a general framework of SC systems under semantic jamming attacks and presents the details of the proposed APG.
Then, Section \ref{ROME} discusses the design of the proposed robust model ensembling approach with multi-level perturbation detection.
Section \ref{Analysis} provides a theoretical analysis of the robustness of ROME and demonstrates its capacity for adaptive robustness.
Simulation results are provided in Section \ref{Simulation} and Section \ref{Conclusion} concludes the whole paper.

\textit{Notations:}
Scalars, vectors, and matrices are respectively denoted by lower case, boldface lower case, and boldface upper case letters.
For a matrix $\mathbf{A}$, $\mathbf{A}^T$ is its transpose, and $||\mathbf{A}||_p$ is a vector containing the $\ell_p$-norm values for each row of the matrix.
For a vector $\mathbf{a}$, $\mathbf{a}^T$ and $||\mathbf{a}||_p$ are its transpose and $\ell_p$-norm, respectively.
Finally, $\mathbb{C}^{m\times n}(\mathbb{R}^{m\times n})$ is the space of $m\times n$ complex (real) matrices.
{Most of the key notations are listed in Table \ref{Notation}.}

\begin{table}[t]\scriptsize
	\centering
	\caption{Definitions of Notations}  
	\label{Notation}
	\begin{tabular}{l|l}
		\toprule
		Notations & Definitions \\
		\midrule
		\midrule
		$\mathbf{s}, \mathbf{z}$ & input image, input random trigger vector \\
		$\mathbf{x}, \Delta\mathbf{x}$ & channel-input symbols and jamming signal\\
		$l, k$ & size of input image and number of channel-input symbols\\
		$\mathbf{y}, \tilde{\mathbf{y}}, \hat{\mathbf{y}}$ & clean, perturbed symbols, and their versatile expression\\
		$\mathbf{f}, \mathbf{f}_k$ & intermediate feature tensor and the $k$-th feature within it\\
		$\mathrm{c}, \mathrm{c}^*, \mathrm{c}_{\mathrm{t}}$ & image class type, inferred class, and true class\\
		$\mathcal{F}_{\bm{\theta}}, \mathcal{G}_{\bm{\phi}}$ & semantic encoder and classifier\\
		$\mathcal{A}_{\bm{\alpha}}, \mathcal{D}_{\bm{\beta}}$ &perturbation generator and detector\\
		$\bm{\theta}, \bm{\phi}, \bm{\alpha}, \bm{\beta}$ & trainable parameters of corresponding DNN models\\
		$\mathcal{P}, \mathcal{S}, \mathcal{M}$ & power normalization, softmax, and model ensembling operators\\
		$\mathcal{G}_{\bm{\phi}_i}, \bm{\phi}_i, \mathcal{G}_i$ & $i$-th base classifier, its parameters, and its simplified expression\\
		$\mathbf{p}_i, \mathbf{p}_{\mathrm{d}}, \mathbf{p_E}$ & prediction of the $i$-th base classifier, detector, and ROME\\
		$\mathbf{P}$ & concatenated prediction map\\
		$\mathrm{h}, \mathrm{h_a}, \mathbf{n}$ & legitimate channel, jamming channel, and AWGN\\
		$\mathbf{y'}, \mathrm{h'_a}$ & target signal and jamming channel generated by the attacker\\
		$N$ & number of base classifiers\\
		$V, L_v$ & pre-defined power level set and the $v$-th power level\\
		$\mathcal{L}_{\mathrm{base}}, \mathcal{L}_d, \mathcal{L}_a$ & loss of base classifiers, detector, and perturbation generator\\
		$\tilde{\mathbf{F}},\mathbf{F},\Delta \mathbf{F},\mathbf{N}$ & tensor form of $\tilde{\mathbf{y}},\mathbf{x},\Delta \mathbf{x}$, and $\mathbf{n}$\\
		$\epsilon, \eta$ & power constraints of semantic jamming signal and AWGN\\
		$\mathbf{V}, \mathbf{E}$ & sets of computational nodes and computational node pairs\\
		$\mathrm{v}_i, \mathrm{u}(i)$ & the $i$-th computational node and the set of its parent nodes\\
		$\mathbf{h}_i, \overline{\mathbf{h}}_i, \underline{\mathbf{h}}_i$ & NN operation of $\mathrm{v}_i$, and its upper and lower linear relaxations\\
		$\overline{\mathbf{W}}_i, \overline{\mathbf{b}}_i, \underline{\mathbf{W}}_i, \underline{\mathbf{b}}_i$ & LiRPA parameters of $\mathrm{v}_i$\\
		$\overline{\mathbf{p}}, \underline{\mathbf{p}}, \overline{\mathbf{p}}_{\mathrm{E}}, \underline{\mathbf{p}}_{\mathrm{E}}$ & upper and lower bounds of a single base classifier and ROME\\
		$\mathbf{B}_k, \mathbf{B}_{\mathrm{E}}$ & distortion bound of the $k$-th base classifier and ROME\\
		$\mathrm{r}_k, \mathrm{r}_{\mathrm{E}}$ & robustness of the $k$-th base classifier and ROME\\
		$C, H, W$ & number, height, and width of intermediate features \\
		$H_{in}, W_{in}$ & height, width of the input image \\
		$K$ & size of convolutional kernels of a convolutional layer \\
		$C_i, C_o$ & number of input and output features of a convolutional layer \\
		$H_o, W_o$ & height and width of output features of a convolutional layer\\
		$H_{out}, W_{out}$ & height and width of encoded semantic features \\
		$D_i, D_o$ & input and output dimension of an FC layer \\
		\bottomrule
	\end{tabular}
\end{table}

\begin{figure}[t]
	\centering
	\includegraphics[width=0.49\textwidth]{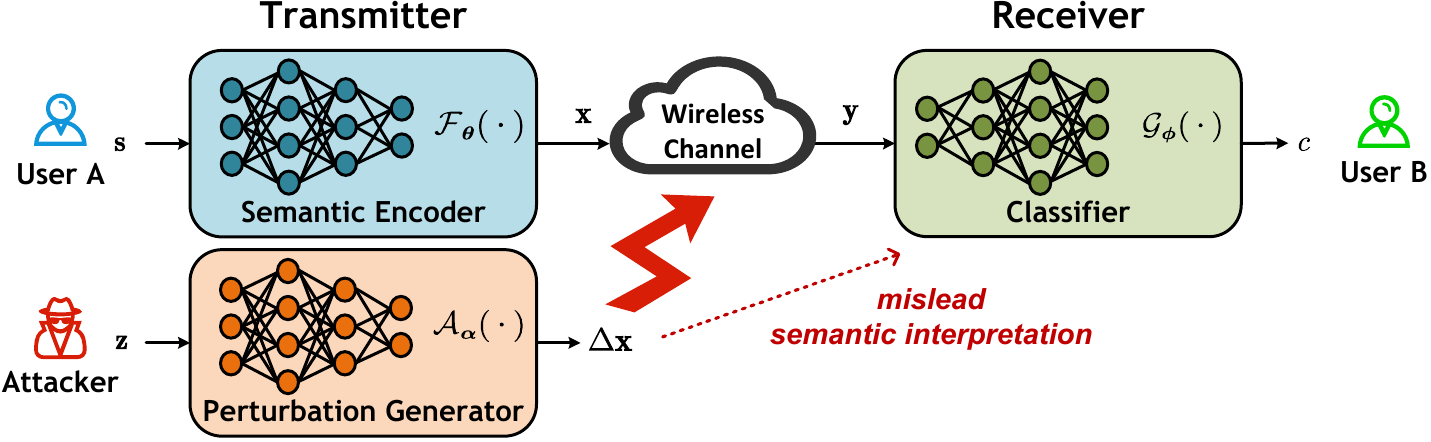}
	\captionsetup{font=footnotesize}
	\caption{The overview of an SC system under semantic jamming attacks.}
	\label{Framework}
\end{figure}

\section{System Model} \label{System}

In this section, we present an overview of an SC system under semantic jamming attacks.

\subsection{SC System}
As shown in Fig. \ref{Framework}, an input image is represented by $\mathbf{s}\in \mathbb{R}^l$, where $l$ is the size of the image.
The transmitter firstly maps $\mathbf{s}$ into a stream of channel-input symbols $\mathbf{x}\in \mathbb{C}^k$, where $k$ is the number of symbols.
Moreover, $\mathbf{x}$ is subject to the average power constraint $P$ at the transmitter, i.e.,
$||\mathbf{x}||^2\leq kP$.
The process is represented as
{
\begin{IEEEeqnarray}{rCl}\label{EncoderFunc}
	\mathbf{x}=\mathcal{P}(\mathcal{F}_{\bm{\theta}}(\mathbf{s})), 
\end{IEEEeqnarray}
where $\bm{\theta}$ denotes the parameter set of the semantic encoder $\mathcal{F}_{\bm{\theta}}(\cdot)$, and $\mathcal{P}(\cdot)$ represents the power normalization process.}
Then, the symbols received at the receiver are given by
{
\begin{IEEEeqnarray}{rCl}\label{ChannelFunc}
 \mathbf{y}=\mathrm{h}\mathbf{x} + \mathbf{n}, 
\end{IEEEeqnarray}
where $\mathrm{h}\in \mathbb{C}$ is the channel realization and $\mathbf{n}\in \mathbb{C}^k$ is additive white Gaussian noise (AWGN) sampled from $\mathcal{CN}(0, \sigma^{2}\mathbf{I})$.}
The receiver leverages $\mathbf{y}$ to infer the category of $\mathbf{s}$.
{In particular, the probability of each class type $\mathrm{c}$ is computed, given by
\begin{IEEEeqnarray}{rCl}
	\mathrm{p}(\mathrm{c}|\mathbf{y})=\mathcal{G}_{\bm{\phi}}(\mathbf{y}), 
\end{IEEEeqnarray}
where $\bm{\phi}$ represents the parameter set of the classifier {$\mathcal{G}_{\bm{\phi}}(\mathbf{y})$}.
Finally, the inferred category of $\mathbf{s}$ is given as
\begin{IEEEeqnarray}{rCl}
	\mathrm{c}^*=\mathop{\arg}\max\limits_{c}\mathrm{p}(\mathrm{c}|\mathbf{y}).
\end{IEEEeqnarray}}\noindent

\subsection{Generation of Power-Variable Semantic Jamming Attacks}
The objective of semantic jamming attacks is to generate and transmit a meticulously crafted imperceptible jamming signal over the wireless channel.
This signal is intended to mislead the semantic interpretation at the receiver without being perceived.
\subsubsection{Problem Formulation}
We denote the semantic jamming signal as $\Delta \mathbf{x}$, and then the perturbed semantic signal can be expressed as
{
\begin{IEEEeqnarray}{rCl}\label{AdvFunc}
	\tilde{\mathbf{y}}=\mathrm{h}\mathbf{x} + \mathrm{h_a}\Delta \mathbf{x} + \mathbf{n},
\end{IEEEeqnarray}
where $\mathrm{h_a}\in \mathbb{C}$ denotes the jamming channel between the attacker and the legitimate receiver.
Then, the generation of semantic jamming attacks can be formulated as the following problem:
\begin{IEEEeqnarray}{lCl}
	& \mathop{\max}\limits_{\Delta\mathbf{x}} \,\, & \mathcal{L}_c\,(\mathcal{G}_{\bm{\phi}}(\tilde{\mathbf{y}}), \mathrm{c_t})\IEEEyesnumber \IEEEyessubnumber\\
	&\text{s.t.} & ||\Delta\mathbf{x}||_2 \leq\epsilon. \IEEEyessubnumber  \label{inputAgnostic}
\end{IEEEeqnarray}}\noindent
Here, $\mathcal{L}_c(\cdot)$ denotes the cross-entropy loss function, $\mathrm{c_t}$ is the true category of the original image $\mathbf{s}$, and $\epsilon$ represents the power constraint of the jamming signal.
{However, it is typically infeasible for the attacker to obtain the knowledge of $\mathbf{y}$ and $\mathrm{h_a}$ at any given time in real-time attacks.
This implies that the generation process of semantic jamming signals should be unaware of the specific target signal $\mathbf{y}$ and the exact jamming channel $\mathrm{h_a}$.} 
\subsubsection{DNN-based Perturbation Generation}
{Specifically, we assume that the attacker possesses complete knowledge regarding the target model, as well as the channel statistics of $\mathrm{h}$ and $\mathrm{h_a}$.}
Building upon this, we design a DNN-based perturbation generator {$\mathcal{A}_{\bm{\alpha}}(\cdot)$ and train} its parameters $\bm{\alpha}$ to generate destructive jamming signals.
Since the attacker {is unaware of the specific} signal to be perturbed, the choice for the attacker is to create a universal jamming signal that can effectively perturb any signal within the receiver's input domain, denoted as $I$.
To achieve this, we randomly sample a trigger vector {$\mathbf{z}\sim \mathcal{N}(0, \mathbf{I})$} as the input of the perturbation generator.
This ensures that the generation process {is unaware of the specific} signal to be perturbed.
{The training process of the perturbation generator can be formulated as follows:
\begin{IEEEeqnarray}{lCl}
	& \underset{\bm{\alpha}}{\max}\,\,\,  &
	\mathbb{E}_{
		\,\begin{subarray}{l}
			\mathbf{z} \sim \mathrm{p}(\mathbf{z}), \\
			\mathbf{y'} \sim \mathrm{p}(\mathbf{y}), \\
			\mathbf{h'_a} \sim \mathrm{p}(\mathbf{h_a})
		\end{subarray}
	} 
	[\mathcal{L}_c(\mathcal{G}_{\bm{\phi}}(\mathbf{y'}+\mathrm{h'_a}\mathcal{A}_{\bm{\alpha}}(\mathbf{z})), \mathrm{c_t})] \IEEEnonumber\\
	& \text{s.t.} & ||\mathcal{A}_{\bm{\alpha}}(\mathbf{z})||_2 \leq\epsilon.   \IEEEyesnumber
\end{IEEEeqnarray}
Here, both $\mathbf{y'}$ and $\mathrm{h'_a}$ are generated based on the attacker's knowledge of the target model and the channel statistics.}
\subsubsection{Adjustable Perturbation Generator}

In real communication scenarios, the power of jamming signals can vary due to the attacker's transmit power.
Therefore, we extensively consider a more challenging case where the attacker is able to generate power-variable jamming signals.

\begin{figure}[t]
	\centering
	\includegraphics[width=0.4\textwidth]{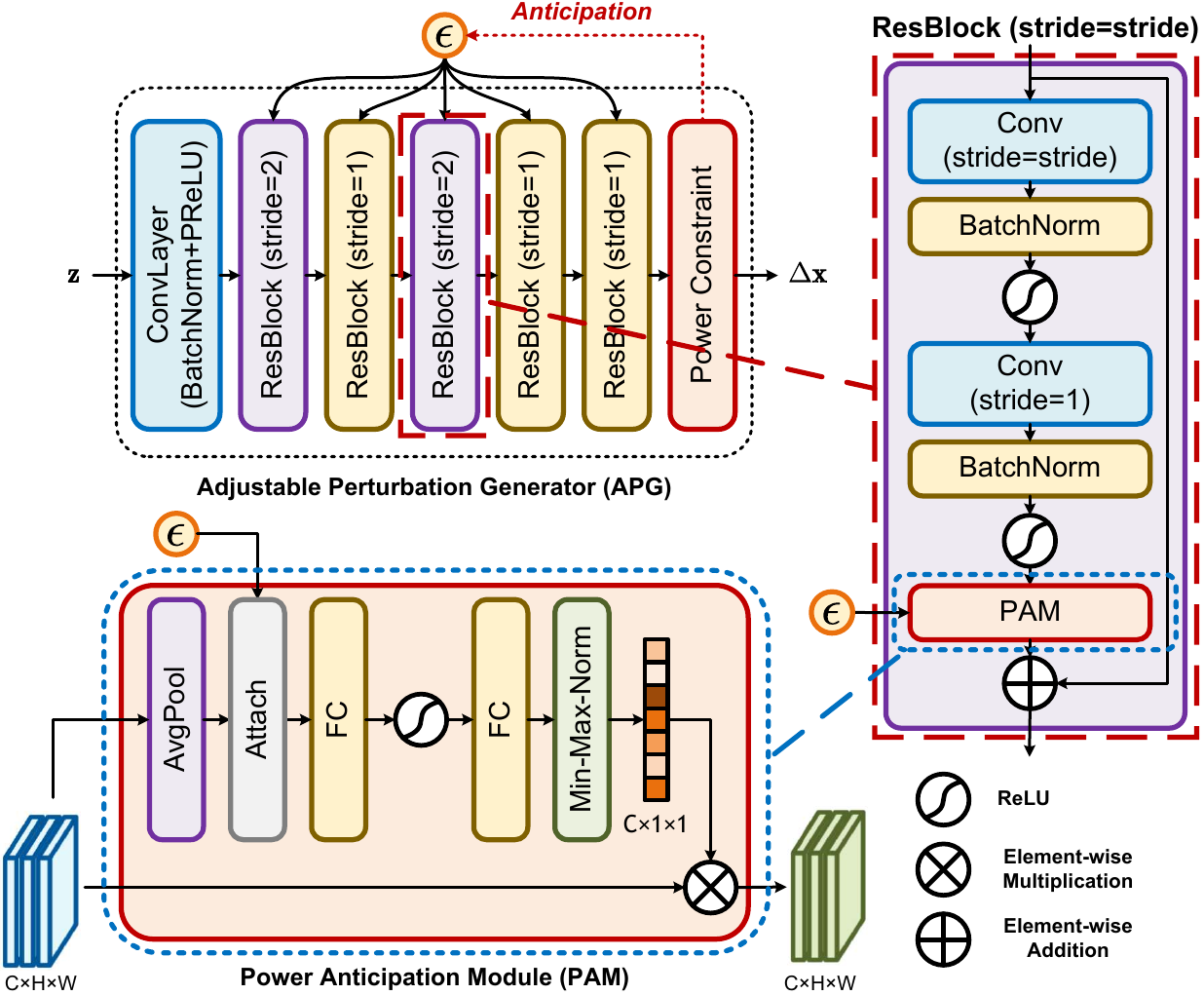}
	\captionsetup{font=footnotesize}
	\caption{The architecture of APG with PAM.}
	\label{APGStructure}
\end{figure}

To this end, we further develop an attack model named APG.
The APG is designed to produce destructive jamming signals that span a broad spectrum of power levels.
As shown in Fig. \ref{APGStructure}, the key idea of the APG is to inform the generator about the power constraint $\epsilon$ ahead of the final power constraining step.
Particularly, we introduce a novel module called the {PAM}.
This module serves as a proactive calibrator, enabling the generator to have anticipation about the upcoming power constraint throughout the perturbation generation process. 
A group of PAMs are inserted into the residual blocks of APG to calibrate its generation process.
Specifically, the inputs of PAM (i.e., the intermediate features) are denoted as $\mathbf{f}\in\mathbb{R}^{C\times H\times W}$, where $C$ is the number of features and $H\times W$ is the shape of each feature map.
These features are first processed by global average pooling and turned into a vector $\mathbf{a}=[\mathrm{a}_1, \mathrm{a}_2, ..., \mathrm{a}_c]^T$, which contains the global information of the features.
Considering the $k$-th feature $\mathbf{f}_k$, this process can be expressed as
\begin{IEEEeqnarray}{rCl}
	\mathrm{a}_k = \frac{1}{HW}\sum_{i=1}^{H}\sum_{j=1}^{W}\mathrm{f}_{k,ij},
\end{IEEEeqnarray}
where $\mathrm{f}_{k,ij}$ is the element in the $i$-th row and the $j$-th column of $\mathbf{f}_k$.
Then, the power constraint $\epsilon$ is attached to the information vector $\mathbf{a}$ as side information.
The extended information vector is represented as $\hat{\mathbf{a}}=[\epsilon, \mathrm{a}_1, \mathrm{a}_2, ..., \mathrm{a}_c]^T$.
Subsequently, $\hat{\mathbf{a}}$ is processed by a couple of fully-connected (FC) layers to produce a calibration vector, denoted as $\tilde{\mathbf{a}}=[\tilde{\mathrm{a}}_1, \tilde{\mathrm{a}}_2, ..., \tilde{\mathrm{a}}_c]^T$, which is then scaled to the interval $[0,1]$ through min-max normalization.
Finally, the calibration vector is utilized to calibrate the input features $\mathbf{f}$.
We {tune} the parameters of our APG together with the inserted PAMs through end-to-end training.

\section{Robust Model Ensembling}\label{ROME}
In general, DNN-based SC systems face an inherent trade-off between robustness and generalization ability.
To effectively counter real-world semantic jamming attacks of varying jamming power levels, it is crucial for the communication system to adaptively balance this trade-off.
Given that a single model only possesses fixed robustness and corresponding generalization ability\footnote{Since robustness is inherently linked to generalization ability, we have simplified phrases like ``robustness and corresponding generalization ability'' to simply ``robustness'' in the following text for conciseness.}, we propose to achieve this adaptive balance by ensembling a collection of base classifiers with different degrees of robustness, thereby covering a broad spectrum of system robustness.
To this end, we propose a novel defense approach named ROME.
\subsection{Overview of ROME}

\begin{figure}[t]
	\centering
	\includegraphics[width=0.4\textwidth]{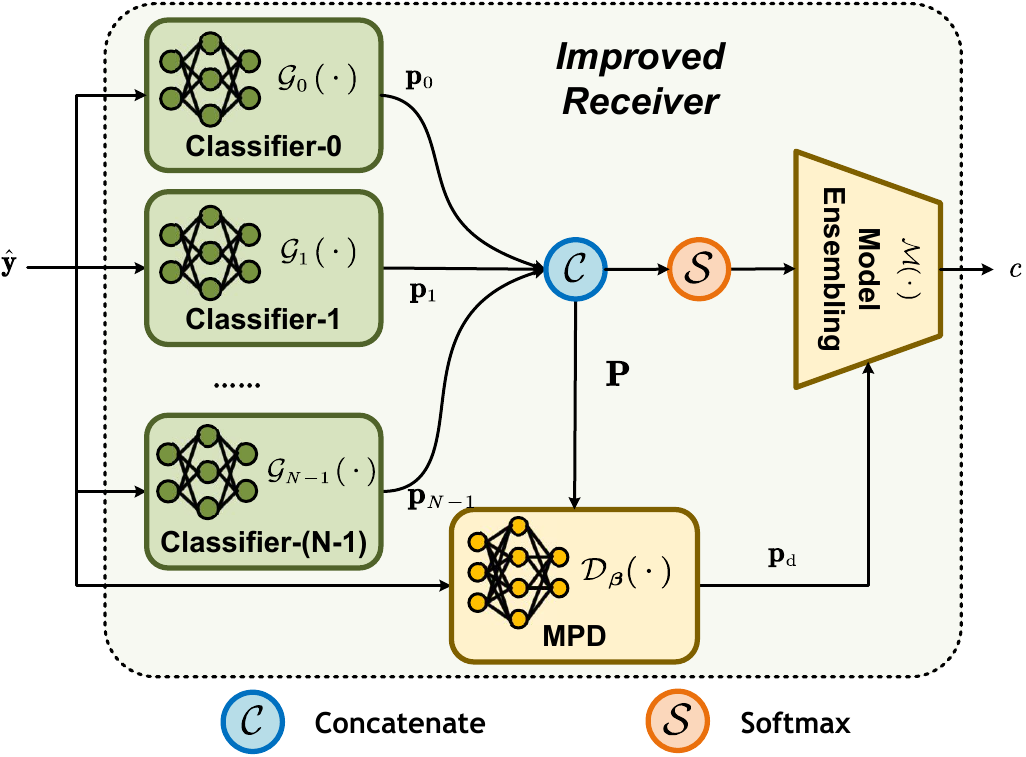}
	\captionsetup{font=footnotesize}
	\caption{The architecture of the receiver with model ensembling.}
	\label{ModelEnsembling}
\end{figure}
	
As illustrated in Fig. \ref{ModelEnsembling}, the ROME approach enhances the receiver by incorporating multiple base classifiers with the proposed MPD.
Each base classifier performs classification tasks independently and contributes to the final decision of the receiver.
In particular, we denote the prediction of the $i$-th classifier as a probability vector, given by
{\begin{IEEEeqnarray}{rCl}
	\mathbf{p}_i=\mathbf{p}_i(\hat{\mathbf{y}}),
\end{IEEEeqnarray}}\noindent
where $\hat{\mathbf{y}}$ serves as a versatile representation of the received signal, encompassing both the clean sample $\mathbf{y}$ as well as the adversarial sample $\tilde{\mathbf{y}}$.
{Then, a prediction map is formed by concatenating the probability vectors, given as
\begin{IEEEeqnarray}{rCl}
	\mathbf{P} = [\mathbf{p}_0,\mathbf{p}_1,\mathbf{p}_2,...,\mathbf{p}_{N-1}]^T.
\end{IEEEeqnarray}
The probability vectors in the prediction map are then normalized to the interval $[0,1]$ using a softmax operator.}

Meanwhile, the MPD continuously detects semantic jamming attacks to provide defense guidance.
It takes the { prediction map $\mathbf{P}$} and the received signal $\hat{\mathbf{y}}$ as input and produces a detection result, denoted by a probability vector $\mathbf{p}_{\mathrm{d}}$.
This probability vector $\mathbf{p}_{\mathrm{d}}$ is then utilized by the model ensembling module to determine the weight of each classifier's prediction and derive the final decision, given by
{\begin{IEEEeqnarray}{rCl}
	\mathbf{p}_{\mathrm{E}} & = & \mathcal{M}(\mathbf{P}, \mathbf{p}_{\mathrm{d}}) = \mathcal{S}(\mathbf{p}_{\mathrm{d}})^T\mathbf{P} = \sum_{i=0}^{N-1}\mathcal{S}(\mathrm{p}_{\mathbf{d}}^{(i)})\mathbf{p}_i \IEEEnonumber*\\
	\mathrm{c}^* & = & \mathop{\arg}\max\limits_{c}\mathrm{p}_{\mathrm{E}}(c|\hat{\mathbf{y}}), \IEEEyesnumber
\end{IEEEeqnarray}
where $\mathcal{M}(\cdot)$ represents the model ensembling module, $\mathcal{S}(\cdot)$ denotes the softmax operator, and $\mathrm{p}_{\mathrm{d}}^{(i)}$ denotes the $i$-th element in the vector $\mathbf{p}_{\mathrm{d}}$.}
The model ensembling module operates as a regulator, which leverages the detection results from MPD to determine the contribution of each base classifier.
In this way, the system's overall robustness and generalization ability can be adaptively balanced, as theoretically demonstrated in Section \ref{Analysis}.
\subsection{Base Classifiers}
To obtain base classifiers with different degrees of robustness, we leverage the AT technique.
The details are presented as follows.
\subsubsection{Adversarial Training}
The main idea of AT involves incorporating adversarial attacks during the training phase, enabling the trained models to behave better when facing adversarial attacks.
Specifically, the objective of AT can be regarded as minimizing the worst-case loss of the system, which can be formulated as
{\begin{IEEEeqnarray}{lCl}\label{advTrain}
	& \mathop{\min}\limits_{\bm{\phi}} &
	\sum_{\mathbf{y}\in I}\mathop{\max}\limits_{\Delta\mathbf{x}} \,\,\mathcal{L}_c\,(\mathcal{G}_{\bm{\phi}}(\tilde{\mathbf{y}}), \mathrm{c_t})
	\IEEEnonumber\\
	&\text{s.t.} & ||\Delta\mathbf{x}||_2 \leq\epsilon \IEEEyesnumber.
\end{IEEEeqnarray}
The inner process expects to find an effective jamming signal that maximizes the system loss.
Then, the outer process aims to minimize the worst-case system loss by training the parameters of the classifier.}
\subsubsection{Acquisition of Base Classifiers}
The classifiers trained with stronger attacks are known to exhibit higher robustness \cite{Madry2018towards, Addepalli2022scaling}.
Therefore, the robustness of the obtained base classifiers can be controlled by adjusting the jamming power constraint $\epsilon$ during the AT process.
The detailed procedures are presented as follows:
\begin{itemize}
	\item [(i)] Train a standard classifier via end-to-end learning without any semantic jamming attacks.
	Denote this trained classifier as {$\mathcal{G}_{\bm{\phi}_0}(\cdot)$}, where $\bm{\phi}_0$ represents its parameter set.
	\item [(ii)] {Train the parameters of $\mathcal{G}_{\bm{\phi}_0}(\cdot)$ under semantic jamming attacks based on (\ref{advTrain}), with the jamming power constraint $\epsilon$ set to $[\epsilon_0, \epsilon_1]$.
	This updated classifier is denoted as $\mathcal{G}_{\bm{\phi}_1}(\cdot)$.}
	\item [(iii)] Repeat step (ii) to obtain additional base classifiers {$\mathcal{G}_{\bm{\phi}_i}(\cdot)$} by adjusting $\epsilon$ to $[\epsilon_{i-1},\epsilon_i]$ ($i = 1,...,N-1$), where $\epsilon_0<\epsilon_1<...<\epsilon_{N-1}$.
\end{itemize}
In this manner, a collection of base classifiers are obtained, each possessing a distinct degree of robustness.
For conciseness, we refer to each base classifier {$\mathcal{G}_{\bm{\phi}_i}(\cdot)$} as $\mathcal{G}_i$ in the following text.
\subsection{Multi-Level Perturbation Detector}
To accommodate real-world semantic jamming attacks with varying power levels, the system needs to dynamically balance the trade-off between robustness and generalization ability.
Therefore, with the base classifiers in place, we still need to consider two key questions:

\indent\textbf{Q1:} When should we balance the trade-off? 
 
\indent\textbf{Q2:} To what extent should we balance the trade-off?\\
\noindent
Notably, an attacker tends to attack semantic models during legitimate users' communication process rather than transmitting jamming signals continuously.
Therefore, employing a robust model becomes {less effective} and unnecessary when there is no attacker in the wireless environment, a scenario that exists most of the time.
Moreover, the jamming power can be variable due to the attacker's transmit power, which means that the {ideal balance} should be tailored to the current jamming power.
Therefore, the answer would be the design of an auxiliary detector, which possesses the capability to identify the presence of jamming signals (addresses \textbf{Q1}) and to measure their power levels (addresses \textbf{Q2}).
To this end, we develop the MPD, whose details are presented as follows.

\subsubsection{Perturbation Power Level Definition}
Considering the continuous nature of jamming power and the finite precision of detection, we commence by categorizing jamming power into $N$ discrete levels and implement our MPD as a DNN-based $N$-class classifier.
Particularly, the power level of a jamming signal is denoted as $L_{v}\in V$, where $V=\{L_0,L_1,L_2,...,L_{N-1}\}$ denotes the set of all the predefined levels, with a special $L_0$ representing the no-attack case.
We assume the power of the jamming signals falls within the range $[\epsilon_0,\epsilon_{N-1}]$.
Then for each level $L_i(i>0)$, we have $\epsilon\in[\epsilon_{i-1},\epsilon_i]$, where $\epsilon_0<\epsilon_1<...<\epsilon_{N-1}$.

\begin{figure}[t]
	\centering
	\includegraphics[width=0.4\textwidth]{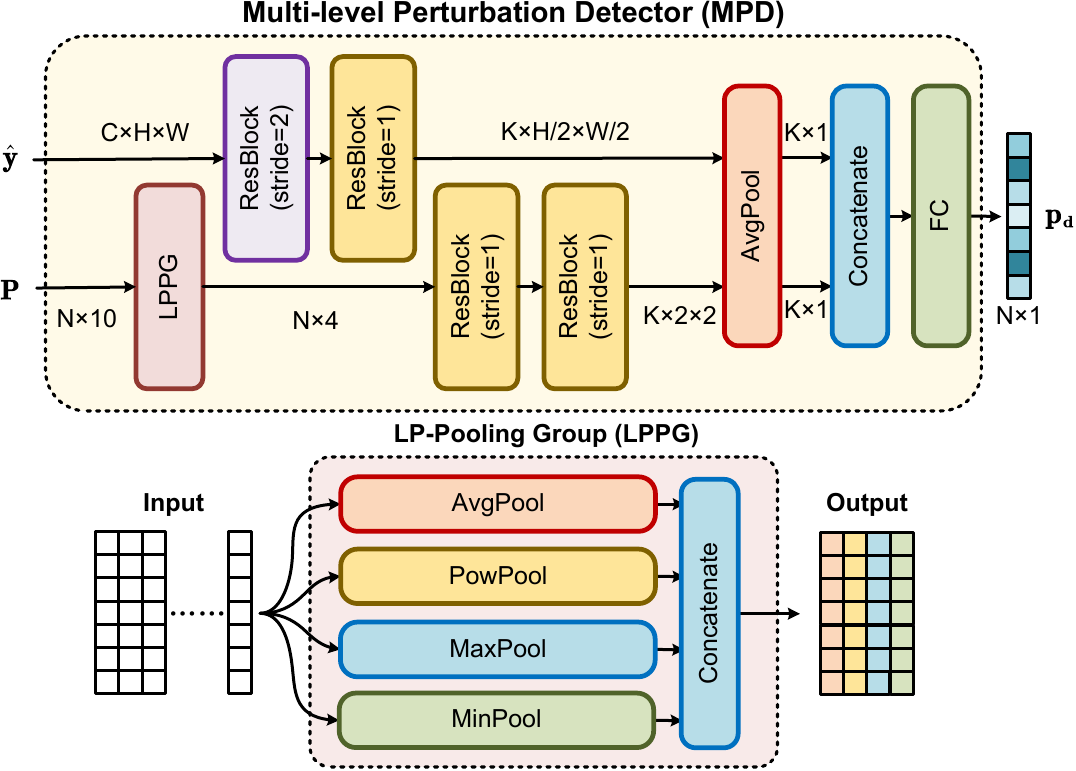}
	\captionsetup{font=footnotesize}
	\caption{The architecture of MPD.}
	\label{MPDStructure}
\end{figure}

\begin{remark}
	The {ideal power level definition} should be tailored for the base classifiers.
	For instance, if the base classifier $\mathcal{G}_i$ outperforms other base classifiers when $\epsilon\in[\epsilon_a,\epsilon_b]$, then the {ideal power level} $L_i$ should be defined as $\epsilon\in[\epsilon_a,\epsilon_b]$.
	In this manner, the model ensembling module will assign the highest weight to $\mathcal{G}_i$ when $\epsilon\in[\epsilon_a,\epsilon_b]$, which ensures the { ideal system performance}.
\end{remark}
\subsubsection{Architecture Design}
As shown in Fig. \ref{MPDStructure}, the inputs to the MPD consist of two key components: the received signal $\hat{\mathbf{y}}$ and the {prediction map $\mathbf{P}$}.
The received signal preserves the most primitive information of both legitimate and jamming signals, serving as the primary reference for the MPD.
On the other hand, the raw prediction map directly reflects the behaviors of the base classifiers, serving as an auxiliary reference.
Since the base classifiers' behaviors vary with changes in jamming power, the MPD can feasibly infer the jamming power level indirectly by observing these behaviors.
These two streams of information are processed in parallel and merged in the end to produce a final detection result, which is the probability of each jamming power level, given as
{\begin{IEEEeqnarray}{rCl}\label{Pd}
	\mathbf{p}_{\mathrm{d}} & = & \mathcal{D}_{\bm{\beta}}(\hat{\mathbf{y}},\mathbf{P}) \IEEEnonumber \\
	& = & [\mathrm{p}(L_0|\hat{\mathbf{y}},\mathbf{P}),\mathrm{p}(L_1|\hat{\mathbf{y}},\mathbf{P}),...,\mathrm{p}(L_{N-1}|\hat{\mathbf{y}},\mathbf{P})]^T, \IEEEeqnarraynumspace
\end{IEEEeqnarray}
where $\mathcal{D}_{\bm{\beta}}(\cdot)$ and $\bm{\beta}$ denote the MPD and its parameter set, respectively.}

Additionally, we introduce a novel feature extraction module called the $\ell_p$-pooling group (LPPG), designed to enhance the MPD's ability to detect jamming signals and facilitate its training process. 
The key operation in the LPPG module is $\ell_p$-pooling, which is defined as follows:
\begin{IEEEeqnarray}{rCl}
	\ell_p(\mathbf{X}) = (\sum_{\mathrm{x}\in\mathbf{X}}\mathrm{x}^p)^{\frac{1}{p}}.
\end{IEEEeqnarray}
Specifically, $\ell_1(\mathbf{X})$ and $\ell_2(\mathbf{X})$ correspond to average pooling and power-average pooling, respectively.
When $p$ approaches $\infty$, the operations $\ell_{\infty}(\mathbf{X})$ and $-\ell_{\infty}(\mathbf{-X})$ correspond to max pooling and min pooling, respectively.
While $\ell_1(\mathbf{X})$ and $\ell_2(\mathbf{X})$ can extract global information from the input, $\ell_{\infty}(\mathbf{X})$ and $-\ell_{\infty}(\mathbf{-X})$ are used to capture local abnormal values.
Therefore, we construct the LPPG module by combining these four pooling operations, as illustrated in Fig. \ref{MPDStructure}.
Note that the pooling kernels in our LPPG are vector-wise, which means the pooling operations are conducted over each probability vector $\mathbf{p}_i$ in the prediction map {$\mathbf{P}$}.
For each probability vector, the aforementioned four pooling operations are performed in parallel, and the extracted features are concatenated and passed to the subsequent layers.

The LPPG module is inserted in the processing flow of {$\mathbf{P}$} to fully realize its potential.
Note that the raw prediction map {$\mathbf{P}$} is only one step away from the system's output, which indicates the impacts of the jamming signal can be explicitly reflected in {$\mathbf{P}$}.
Hence, the LPPG module can help the MPD better capture the jamming-related features from {$\mathbf{P}$}, thereby enhancing the MPD's detection capability and accelerating network convergence.
Conversely, the raw received signal $\hat{\mathbf{y}}$ is a mixture of the jamming and legitimate signals, where the jamming signal is almost imperceptible.
This renders it challenging to directly extract the jamming-related features from $\hat{\mathbf{y}}$.
Therefore, we construct the processing flow of $\hat{\mathbf{y}}$ using only DNN layers with sufficient depth.
\subsubsection{Training}

\begin{algorithm}[t]
	\begin{small}
		\caption{Training algorithm for MPD} 
		\label{TrainMPD}
		\DontPrintSemicolon
		\SetKwInOut{Input}{Input}
		\SetKwInOut{Output}{Output}
		\KwIn{The training epochs $I$, perturbation level set $V$, dataset $S$, pre-trained {semantic encoder $\mathcal{F}_{\bm{\theta}}(\cdot)$, semantic jamming attack model and MPD $\mathcal{D}_{\bm{\beta}}(\cdot)$.}}
		\KwOut{The {trained} parameter set $\bm{\beta}$.}
		\For{$i\leftarrow 1$ \KwTo $I$}{
			\For{$\mathbf{s}\in S$}{
				Randomly sample a power level $L_v$ from $V$.\\
				Compute $\mathbf{x}$ using {$\mathcal{F}_{\bm{\theta}}(\cdot)$}.\\
				\uIf {$L_v\neq L_0$} {
					Randomly sample $\epsilon$ from the range $[\epsilon_{v-1},\epsilon_v]$.\\
					Generate $\Delta \mathbf{x}$ based on $\epsilon$.\\
					Compute $\hat{\mathbf{y}}$ based on (\ref{AdvFunc}).
				}
				\Else {
					Compute $\hat{\mathbf{y}}$ based on (\ref{ChannelFunc}).
				}
				Compute the prediction results $\mathbf{p}_{\mathrm{d}}$ based on (\ref{Pd}).\\
				Compute the loss $\mathcal{L}_d$.\\
				Update $\bm{\beta}$.
			}
		}
	\end{small}
\end{algorithm}

We {train} the parameters of our MPD using both attacked signal samples and clean ones.
During training, the jamming power level $L_v$ is randomly sampled from $V$ for each mini-batch.
If the power level is $L_0$, a mini-batch of clean samples will be fed into the MPD.
Otherwise, a mini-batch of adversarial samples will be generated based on the chosen power level and fed into the MPD.
The training details are presented in Algorithm \ref{TrainMPD}, where $\mathcal{L}_d$ represents the loss function of the MPD, given by
\begin{IEEEeqnarray}{rCl}
	\mathcal{L}_d = \mathcal{L}_c(\mathbf{p}_{\mathrm{d}},L_v).
\end{IEEEeqnarray}

\subsection{Case Study}
The attack-and-defense framework between the attacker and legitimate user can be viewed as a semantic communication game.
The dynamics of this game vary depending on the information each party possesses about the other.
From the perspective of the legitimate user, we conduct a comprehensive case study to fully present the efficacy of the proposed scheme.
\subsubsection{Ideal Case}
In the {ideal case}, the attacker is fully exposed to the legitimate user, granting the legitimate user complete knowledge of the APG {$\mathcal{A}_{\bm{\alpha}}(\cdot)$}.
This allows the legitimate to establish a proactive defense.

Specifically, during the AT process, the legitimate user {trains} the base classifiers using adversarial samples generated by {$\mathcal{A}_{\bm{\alpha}}(\cdot)$}, enhancing the base classifiers' performance against this attacker's attacks.
The detection accuracy of the MPD is also improved by training it with {$\mathcal{A}_{\bm{\alpha}}(\cdot)$}.
Moreover, the {ideal power level definition} can be figured out in this scenario, ensuring the best alignment between the MPD and the base classifiers.
\subsubsection{General Case}
In the general case, the attacker has partial knowledge about the legitimate user, such as the standard model $\mathcal{G}_0$.
This situation is referred to as a grey-box attack scenario.
Conversely, the true nature of the attacker remains completely unknown to the legitimate user, forcing the legitimate user to adopt passive defense measures.

During the AT process, an agent attack method is employed instead of the APG {$\mathcal{A}_{\bm{\alpha}}(\cdot)$}.
In particular, we adopt the projected gradient descent (PGD) attack \cite{Madry2018towards} as the agent attack method in our research.
It is essential to note that the MPD is also a DNN-based classifier, which means the attacker can potentially deceive it.
To mitigate this risk, we also employ the PGD-based AT technique to strengthen the MPD's robustness.
Moreover, note that it is infeasible to determine the {ideal power level definition} since the base classifiers' performance under APG attacks is unknown.
An alternative approach is to adopt the power level used during the AT process, which means defining $L_i$ as $\epsilon\in[\epsilon_{i-1},\epsilon_i]$.
\subsubsection{Worst Case}
In the worst case, the attacker has complete knowledge of all models used by the legitimate user, including the MPD and the base classifiers.
Conversely, the legitimate user has no knowledge of the attacker.
In this scenario, the legitimate user's defense strategy remains the same as in the general case, while the attacker's model is strengthened.

Specifically, the loss function of the APG, denoted as $\mathcal{L}_a$, is modified by integrating the losses of the MPD and base classifiers, defined as
{\begin{IEEEeqnarray}{rCl}
	\mathcal{L}_a & = &  -\mathcal{L}_{\mathrm{base}} - \mathcal{L}_d \IEEEnonumber \\
	& = & -\mathcal{L}_c(\sum_{i=0}^{N-1}\mathrm{w}_i\mathcal{G}_i(\mathbf{y'}+\mathrm{h'_a}\mathcal{A}_{\bm{\alpha}}(\mathbf{z})),\mathrm{c_t}) - \mathcal{L}_c(\mathbf{p}_{\mathrm{d}},L_v). \IEEEeqnarraynumspace
\end{IEEEeqnarray}}\noindent
The first term aims to maximize the loss of the base classifiers, while the second term focuses on maximizing the loss of the MPD.
Additionally, the first term employs an inner-ensembling approach, using a single cross-entropy calculation for the ensembled output of the base classifiers.
This design of the first term aims to further strengthen the attack model, which has proven to be more effective than calculating and then ensembling the cross-entropy losses of each base classifier \cite{Dong2018boosting}.

\section{Robustness Analysis}\label{Analysis}
In this section, we further elaborate on the robustness of the proposed ROME approach.
We begin with a brief introduction of the robustness verification problem (RVP) \cite{Katz2017reluplex}.
Following this, we provide a theoretical analysis of the distortion bound of ROME, formulate its robustness, and demonstrate its superiority in accommodating power-variable semantic jamming attacks.
\subsection{Introduction of RVP} \label{rvpIntro}
DNNs are known to be vulnerable to certain perturbed inputs, which can undermine their reliability despite their ability to generalize to unseen data.
To address this issue, RVP has been proposed to certify the robustness of DNNs and provide theoretical guarantees for their behaviors.
Given inputs perturbed within a certain bound, RVP aims to investigate the corresponding dynamics of the DNN's outputs and derive an output distortion bound.
If the outputs are bounded within a small range for all perturbed inputs within the input space, the DNN is considered robust.
Conversely, the DNN is deemed to lack robustness if the outputs fluctuates significantly \cite{He2024rate}.
However, finding the exact distortion bound is an NP-complete problem \cite{Katz2017reluplex}, rendering formal verification computationally intractable, especially for large-scale DNNs.
Recent studies on RVP have made significant breakthroughs in addressing this challenge, deriving non-trivial certified output bounds that are computationally feasible \cite{Zhang2018efficient,Singh2019abstract,Weng2018towards,Wang2018efficient}.
The theoretical techniques employed in these studies are referred to as linear relaxation-based perturbation analysis (LiRPA), which have been unified and generalized to broader machine learning models \cite{Xu2020automatic}.
\subsection{RVP in SC Systems}
We introduce RVP to the framework of SC systems.
We start by considering a scenario where the system has only a single standard classifier $\mathcal{G}$ at the receiver.
\subsubsection{Input Perturbations}
Given an SC system subject to semantic jamming attacks, input perturbations arise from both channel noise and semantic jamming attacks.
{Therefore, we model the perturbed inputs as
\begin{IEEEeqnarray}{rCl}\label{tensormodel}
	\tilde{\mathbf{F}}=\mathrm{h}\mathbf{F} + \mathrm{h_a}\Delta \mathbf{F} + \mathbf{N},
\end{IEEEeqnarray}}\noindent
where $\tilde{\mathbf{F}},\mathbf{F},\Delta \mathbf{F},\mathbf{N} \in \mathbb{R}^{c\times h\times w}$ are tensor-form representations of $\tilde{\mathbf{y}},\mathbf{x},\Delta\mathbf{x},$ and $\mathbf{n}$, respectively, $c$ is the number of semantic features, and $h\times w$ is the size of each feature.
{By employing equalization at the legitimate receiver, the model can be expressed as
\begin{IEEEeqnarray}{rCl}
	\frac{1}{\mathrm{h}}\tilde{\mathbf{F}}=\mathbf{F} + \frac{\mathrm{h_a}}{\mathrm{h}}\Delta \mathbf{F} + \frac{1}{\mathrm{h}}\mathbf{N}.
\end{IEEEeqnarray}
For simplicity, we redefine the variables to represent the model as Eq. (\ref{tensormodel}):
\begin{IEEEeqnarray}{rCl}
	\tilde{\mathbf{F}}\triangleq\frac{1}{\mathrm{h}}\tilde{\mathbf{F}},\,\,\Delta \mathbf{F}\triangleq\frac{\mathrm{h_a}}{\mathrm{h}}\Delta \mathbf{F}, \,\,\mathbf{N}\triangleq\frac{1}{\mathrm{h}}\mathbf{N}, 
\end{IEEEeqnarray}
where the redefined $\Delta \mathbf{F}$ and $\mathbf{N}$ follow distinct distributions compared to those in Eq. (\ref{tensormodel}).
By specifying $\epsilon$ and $\eta$ to meet the conditions where $P(||\Delta\mathbf{F}||_p\leq\epsilon)\geq 1-\delta$ and $P(||\mathbf{N}||_p\leq\eta)\geq 1-\delta$, with $\delta$ being a very small positive number, we can consider the perturbations to be approximately subject to the constraints:
\begin{IEEEeqnarray}{rCl}
	||\Delta \mathbf{F}||_p \leq \epsilon, \quad ||\mathbf{N}||_p \leq \eta,
\end{IEEEeqnarray}
where $||\cdot||_p$ denotes the $\ell_p$-norm, $\epsilon$ is positively related to the attacker's jamming power, and $\eta$ is negatively related to the signal-to-noise-ratio (SNR).}
Then, the perturbed inputs $\tilde{\mathbf{F}}$ can be considered within a bounded $\ell_p$-ball centered at $\mathbf{F}$, i.e., $\tilde{\mathbf{F}}\in \mathbb{B}_p(\mathbf{F}, \epsilon+\eta)$, where $\mathbb{B}_p(\mathbf{F}, \epsilon+\eta)=\{\tilde{\mathbf{F}}: ||\tilde{\mathbf{F}}-\mathbf{F}||_p\leq\epsilon+\eta\}$\footnote{According to the triangle inequality, we can readily deduce that $||\tilde{\mathbf{F}}-\mathbf{F}||_p=||\Delta \mathbf{F}+\mathbf{N}||_p\leq ||\Delta \mathbf{F}||_p + ||\mathbf{N}||_p\leq \epsilon+\eta$.}.
\subsubsection{Output Distortion Bound}
To characterize the system's robustness, we aim to derive the distortion bound of the system’s output w.r.t. all the perturbed inputs within the input space $\mathbb{B}_p(\mathbf{F}, \epsilon+\eta)$.
Specifically, our goal is to find the upper and lower bounds of the system's output $\mathbf{p}$, denoted as $\overline{\mathbf{p}}$ and $\underline{\mathbf{p}}$, respectively.
Then, the distortion bound of the system w.r.t. all the perturbed inputs is defined as
\begin{IEEEeqnarray}{rCl} \label{DistortionBound}
	\mathbf{B}\triangleq B(\mathcal{G}, \mathbf{F}, \epsilon+\eta)=\max_{\tilde{\mathbf{F}}}\overline{\mathbf{p}} - \min_{\tilde{\mathbf{F}}}\underline{\mathbf{p}}.
\end{IEEEeqnarray}
As mentioned in Section \ref{rvpIntro}, the distortion bound is negatively related to the system's robustness.
Thus, we establish the connection between them by $||\mathbf{B}||_p\propto\frac{1}{r}$.
However, deriving the exact upper and lower bounds can be computationally demanding due to the non-linear operations in DNNs.
To tackle this issue, we employ the LiRPA technique in \cite{Xu2020automatic}.

\begin{figure}[t]
	\centering
	\includegraphics[width=0.35\textwidth]{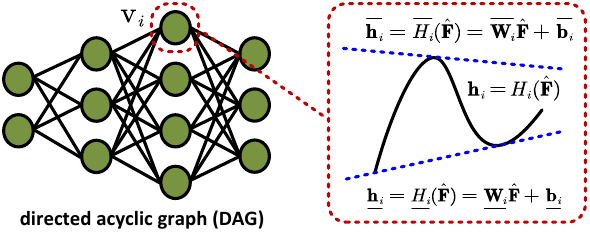}
	\captionsetup{font=footnotesize}
	\caption{Linear bounds of a computation node.}
	\label{LinearRelax}
\end{figure}

As illustrated in Fig. \ref{LinearRelax}, the base classifier $\mathcal{G}$ can be considered as a directed acyclic graph (DAG) $\mathbf{G}_{\mathcal{G}}$ = ($\mathbf{V}$, $\mathbf{E}$), where $\mathbf{V}=\{\mathrm{v}_1,\mathrm{v}_2,...,\mathrm{v}_n\}$ is the set of computation nodes in $\mathcal{G}$ and $\mathbf{E}=\{(\mathrm{v}_i,\mathrm{v}_j)\}$ is the set of node pairs representing that $\mathrm{v}_i$
is the input node of $\mathrm{v}_j$.
Each computation node $\mathrm{v}_i$ represents an NN
operation $\mathbf{h}_i=h_i(\mathrm{u}(i))\in\mathbb{R}^{\mathrm{d}_i}$, where $\mathrm{u}(i)$ is the set of
parent nodes for node $\mathrm{v}_i$.
For conciseness, we represent each NN operation as a function of the input feature tensor $\tilde{\mathbf{F}}$, without explicitly referring to the parent nodes, i.e.,  $\mathbf{h}_i=H_i(\tilde{\mathbf{F}})$.

In the LiRPA framework, we can find two linear bounds of $H_i(\tilde{\mathbf{F}})$ for each node $\mathrm{v}_i$ using linear relaxation, defined as
\begin{IEEEeqnarray}{rrCl}
	& \overline{\mathbf{h}}_i & = & \overline{H_i}(\tilde{\mathbf{F}})=\overline{\mathbf{W}}_i\tilde{\mathbf{F}}+\overline{\mathbf{b}}_i,\IEEEyesnumber \IEEEyessubnumber*\\[-0.425\normalbaselineskip]
	\smash{\left\{
		\IEEEstrut[7\jot]
		\right.} \IEEEnonumber\\[-0.425\normalbaselineskip]
	& \underline{\mathbf{h}}_i & = & \underline{H_i}(\tilde{\mathbf{F}})=\underline{\mathbf{W}}_i\tilde{\mathbf{F}}+\underline{\mathbf{b}}_i,
\end{IEEEeqnarray}
where $\underline{\mathbf{h}}_i \preceq \mathbf{h}_i \preceq \overline{\mathbf{h}}_i$.
$\overline{\mathbf{W}}_i, \underline{\mathbf{W}}_i \in\mathbb{R}^{\mathrm{d}_i\times\mathrm{d}_1}, \overline{\mathbf{b}}_i, \underline{\mathbf{b}}_i \in\mathbb{R}^{\mathrm{d}_i}$ are parameters of the linear bounds.
According to the attributes of DAG, we can propagate the linear bounds of each node to its successor nodes in a forward manner until reaching the output node $\mathrm{v}_n$.
Particularly, for each node $\mathrm{v}_i$, we have a forward propagating function $F_i$, which takes the linear bounds parameters of all its parent nodes $\mathrm{v}_j\in u(i)$ as input, i.e., $(\overline{\mathbf{W}}_i, \overline{\mathbf{b}}_i, \underline{\mathbf{W}}_i, \underline{\mathbf{b}}_i)=F_i(\{(\overline{\mathbf{W}}_j, \overline{\mathbf{b}}_j, \underline{\mathbf{W}}_j, \underline{\mathbf{b}}_j)|\mathrm{v}_j\in u(i)\})$.
Note that the parameters of the input node $\mathrm{v}_1$ are $\overline{\mathbf{W}}_1=\underline{\mathbf{W}}_1=\mathbf{1}, \overline{\mathbf{b}}_1=\underline{\mathbf{b}}_1=\mathbf{0}$.
For each type of DNN operation, such as activation functions and matrix multiplication, the forward propagating function $F_i$ is artificially defined.
Details of these functions can be found in previous studies \cite{Zhang2018efficient,Salman2019convex, Shi2019robustness, Wong2018provable} that have covered various common operations in DNNs.
By propagating the linear bounds of each node, we can yield the bounds of the output node $\mathrm{v}_n$, which are also the bounds of the system:
\begin{IEEEeqnarray}{rrCl}
	& \overline{\mathbf{p}} & = & \overline{H_n}(\tilde{\mathbf{F}}) =  \overline{\mathbf{W}}_n\tilde{\mathbf{F}}+\overline{\mathbf{b}}_n, \IEEEyesnumber \IEEEyessubnumber*\\[-0.425\normalbaselineskip]
	\smash{\left\{
		\IEEEstrut[7\jot]
		\right.} \IEEEnonumber\\[-0.425\normalbaselineskip]
	& \underline{\mathbf{p}} & = & \underline{H_n}(\tilde{\mathbf{F}}) =  \underline{\mathbf{W}}_n\tilde{\mathbf{F}}+\underline{\mathbf{b}}_n.
\end{IEEEeqnarray}

\begin{figure*}[t]
	\centering
	\includegraphics[width=0.8\textwidth]{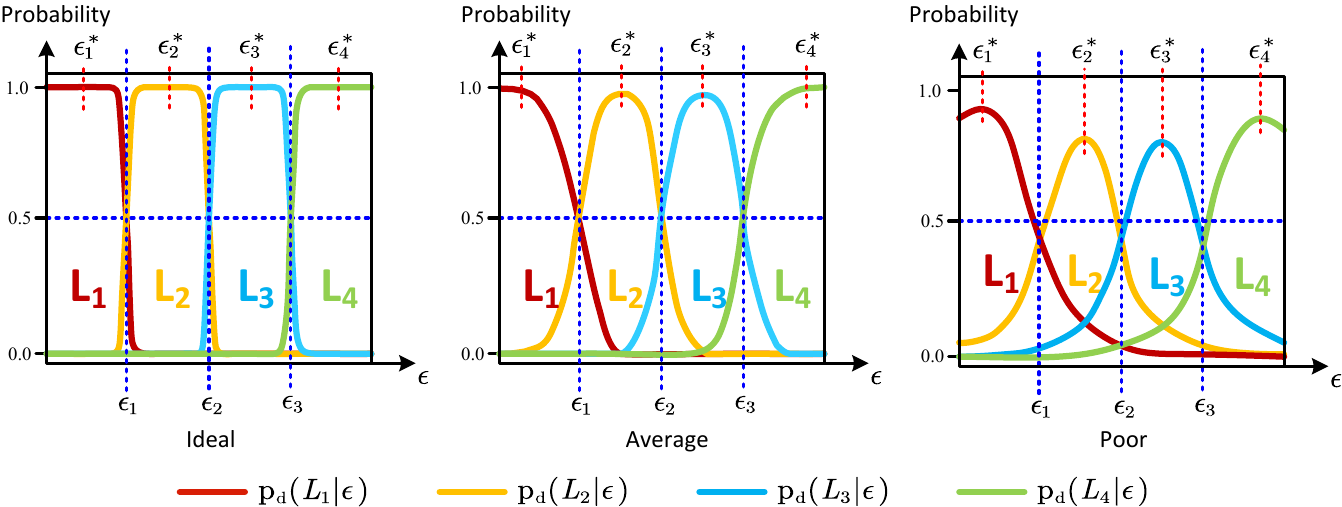}
	\captionsetup{font=footnotesize}
	\caption{Visualization of the MPD's performance. $\epsilon_k^*$ denotes the jamming power where the MPD has the most confidence in $L_k$, $\epsilon_{k-1}$ and $\epsilon_k$ denote the MPD's left and right decision boundaries of $L_k$. The wider the ``eyes'' open, the more accurate the MPD is, leading to better system performance.}
	\label{EyeDiagram}
\end{figure*}

Next, we compute the maximum upper bound and minimum lower bound w.r.t. all the perturbed inputs $\tilde{\mathbf{F}}$, given as
\begin{IEEEeqnarray}{rCl}
	\max_{\tilde{\mathbf{F}}}\overline{\mathbf{p}} & = & \max_{\tilde{\mathbf{F}}}\overline{\mathbf{W}}_n\tilde{\mathbf{F}}+\overline{\mathbf{b}}_n \IEEEnonumber \\
	& = & (\epsilon+\eta)||\overline{\mathbf{W}}_n||_q+\overline{\mathbf{W}}_n\mathbf{F}+\overline{\mathbf{b}}_n \IEEEyesnumber \IEEEyessubnumber,\\
	\min_{\tilde{\mathbf{F}}}\underline{\mathbf{p}} & = & \min_{\tilde{\mathbf{F}}}\underline{\mathbf{W}}_n\tilde{\mathbf{F}}+\underline{\mathbf{b}}_n \IEEEnonumber \\
	& = & -(\epsilon+\eta)||\underline{\mathbf{W}}_n||_q+\underline{\mathbf{W}}_n\mathbf{F}+\underline{\mathbf{b}}_n \IEEEyesnumber \IEEEyessubnumber,
\end{IEEEeqnarray}
where $||\cdot||_q$ is the dual norm of the $\ell_p$-norm (i.e., $\frac{1}{p}+\frac{1}{q}=1$) for each row in the matrix, and the result is a vector.
The detailed derivation is presented in Appendix A.
Finally, we can derive the distortion bound of the system based on (\ref{DistortionBound}), given as
\begin{IEEEeqnarray}{rCl}\label{DistortionBound1}
	\mathbf{B} & = & (\epsilon+\eta)(||\overline{\mathbf{W}}_n||_q+||\underline{\mathbf{W}}_n||_q)+(\overline{\mathbf{W}}_n-\underline{\mathbf{W}}_n)\mathbf{F} \IEEEnonumber \\
	& & +\overline{\mathbf{b}}_n-\underline{\mathbf{b}}_n \IEEEyesnumber.
\end{IEEEeqnarray}

{
The above analysis mainly focuses on the impact of perturbations under the assumption that $\mathbf{F}$ remains constant.
Building upon this foundation, we now delve into the impact of $\mathbf{F}$.
Considering that $\mathbf{F}$ represents the tensor-form of the channel-input symbols $\mathbf{x}$, it is constrained by the transmit power constraint, specifically $||\mathbf{F}||_2 \leq \sqrt{kP}$.
Drawing on the methodology used to derive the maximum upper bound in Appendix A, we can derive the following:
\begin{IEEEeqnarray}{rCl}\label{DistortionBound2}
	\max_{\mathbf{F}}\mathbf{B} & = & (\epsilon+\eta)(||\overline{\mathbf{W}}_n||_q+||\underline{\mathbf{W}}_n||_q) \IEEEnonumber\\
	& & +\sqrt{kP}||\overline{\mathbf{W}}_n-\underline{\mathbf{W}}_n||_2
	+\overline{\mathbf{b}}_n-\underline{\mathbf{b}}_n\triangleq\mathbf{B}.
\end{IEEEeqnarray}
Given the same perturbation constraint $(\epsilon+\eta)$ and transmit power constraint $\sqrt{kP}$, it becomes evident that a classifier with larger LiRPA parameters will exhibit a larger distortion bound, indicating reduced robustness.
Furthermore, it is important to recognize that the robustness of a model is an intrinsic characteristic, independent of the perturbations and inputs.
Thus, we define the robustness of a classifier based on its LiRPA parameters as follows:
\begin{IEEEeqnarray}{rCl}
	r & = & - (||||\overline{\mathbf{W}}_n||_q+||\underline{\mathbf{W}}_n||_q||_p + ||||\overline{\mathbf{W}}_n-\underline{\mathbf{W}}_n||_2||_p \IEEEnonumber \\
	& & + ||\overline{\mathbf{b}}_n-\underline{\mathbf{b}}_n||_p). \IEEEyesnumber
\end{IEEEeqnarray}}\noindent
\subsection{Robustness of ROME}
Next, we extend the above analysis to the framework of ROME, where the system is equipped with $N$ base classifiers and an MPD.
{Considering the classifier $\mathcal{G}_k$, its distortion bound is represented as
\begin{IEEEeqnarray}{rCl}
	\mathbf{B}_k & = & (\epsilon+\eta)(||\overline{\mathbf{W}}^{(k)}_n||_q+||\underline{\mathbf{W}}^{(k)}_n||_q) \IEEEnonumber* \\
	& & +\sqrt{kP}||\overline{\mathbf{W}}^{(k)}_n-\underline{\mathbf{W}}^{(k)}_n||_2 + \overline{\mathbf{b}}^{(k)}_n-\underline{\mathbf{b}}^{(k)}_n, \IEEEyesnumber
\end{IEEEeqnarray}
and the corresponding robustness is defined as
\begin{IEEEeqnarray}{rCl}
	r_k & = & -(||||\overline{\mathbf{W}}_n^{(k)}||_q+||\underline{\mathbf{W}}_n^{(k)}||_q||_p + ||||\overline{\mathbf{W}}_n^{(k)}-\underline{\mathbf{W}}_n^{(k)}||_2||_p \IEEEnonumber \\
	& & + ||\overline{\mathbf{b}}_n^{(k)}-\underline{\mathbf{b}}_n^{(k)}||_p). \IEEEyesnumber
\end{IEEEeqnarray}}\noindent
Without loss of generality, we assume that the base classifiers satisfy $\mathbf{B}_0 \succ \mathbf{B}_1 \succ ... \succ \mathbf{B}_{N-1}$ and $r_0<r_1<...<r_{N-1}$.

Additionally, we consider the MPD's output $\mathbf{p}_{\mathrm{d}}$ as a function of the jamming power $\epsilon$, given as
\begin{equation}
	\mathbf{p}_\mathrm{d}=[\mathrm{p_d}(L_0|\epsilon),\mathrm{p_d}(L_1|\epsilon),...,\mathrm{p_d}(L_{N-1}|\epsilon)]^T.
\end{equation}
To facilitate the subsequent analysis, we assume that the MPD satisfies the following properties:
\begin{assumption}\label{mpdAssume}
	(\textbf{Properties of the MPD})
	\begin{itemize}
		\item [(1)] $\sum_{k=0}^{N-1}\mathrm{p}_{\mathrm{d}}(L_k|\epsilon)=1,\forall \epsilon \in \mathbb{R}$.
		\item[(2)] $\nabla^2_{\epsilon}\mathrm{p}_{\mathrm{d}}(L_k|\epsilon)\leq 0, \forall k\in \{0,...,N-1\}$.
		\item[(3)] $\mathrm{p}_{\mathrm{d}}(L_0|\epsilon_0^*) >> \mathrm{p}_{\mathrm{d}}(L_1|\epsilon_0^*)$,
		
		$\mathrm{p}_{\mathrm{d}}(L_{N-1}|\epsilon_{N-1}^*) >> \mathrm{p}_{\mathrm{d}}(L_{N-2}|\epsilon_{N-1}^*)$,
		
		$\mathrm{p}_{\mathrm{d}}(L_k|\epsilon_k^*) >> \mathrm{p}_{\mathrm{d}}(L_{k-1}|\epsilon_k^*) + \mathrm{p}_{\mathrm{d}}(L_{k+1}|\epsilon_k^*)$,
		
		$\forall k\in \{1,...,N-2\}$, where $\nabla_{\epsilon}\mathrm{p}_{\mathrm{d}}(L_k|\epsilon=\epsilon_k^*) = 0$.
		\item[(4)] $\mathrm{p}_{\mathrm{d}}(L_k|\epsilon_{k-1})\rightarrow 0.5,\mathrm{p}_{\mathrm{d}}(L_k|\epsilon_k)\rightarrow 0.5$,
		
		$\forall k\in \{1,...,N-1\}$.
	\end{itemize}
	Here, $\epsilon_k^*$ denotes the jamming power where the MPD has the most confidence in $L_k$, $\epsilon_{k-1}$ and $\epsilon_k$ denote the MPD's left and right decision boundaries of $L_k$, respectively.
\end{assumption}
The explanations for the properties outlined in Assumption \ref{mpdAssume} are as follows.
Property (1) states that the MPD serves as an $N$-class classifier.
Property (2) and (3) imply that the MPD's confidence in $L_i$ reaches its maximum when the jamming power $\epsilon$ is within the level $L_i$, and this confidence decreases as $\epsilon$ deviates from the range of $L_i$.
Property (4) characterizes the MPD's behaviors at decision boundaries.
Fig. \ref{EyeDiagram} is a visualization of these properties, which illustrates the MPD's detection confidence at different jamming power levels, reflecting its performance.
The wider the ``eyes'' open, the more accurate the MPD is, leading to better system performance.

Considering the fact that the output of ROME is a weighted sum of the base classifiers' outputs, i.e., $\mathbf{p}_{\mathrm{E}}=\sum_{i=0}^{N-1}\mathrm{p}_{\mathrm{d}}(L_i|\epsilon)\mathbf{p}_i$, we can easily derive the linear bounds of ROME as
\begin{IEEEeqnarray}{rrCl}
	& \overline{\mathbf{p}}_{\mathrm{E}} & = & \sum_{i=0}^{N-1} \mathrm{p}_\mathrm{d}(L_i|\epsilon)\overline{\mathbf{p}}_i \IEEEyesnumber \IEEEyessubnumber*\\[+0.125\normalbaselineskip]
	\smash{\left\{
		\IEEEstrut[9\jot]
		\right.} \IEEEnonumber\\[-0.325\normalbaselineskip]
	& \underline{\mathbf{p}}_{\mathrm{E}} & = & \sum_{i=0}^{N-1} \mathrm{p}_\mathrm{d}(L_i|\epsilon)\underline{\mathbf{p}}_i,
\end{IEEEeqnarray}
where $\overline{\mathbf{p}}_{\mathrm{E}}$ and $\underline{\mathbf{p}}_{\mathrm{E}}$ denote the upper and lower bounds of ROME, respectively.
Next, we proceed with our analysis in a special-to-general manner as follows and the detailed derivation can be found in Appendix B.
\begin{itemize}
	\item \textbf{Confident Case}.
	This is the case when $\epsilon=\epsilon_k^*$.
	According to Assumption \ref{mpdAssume}, we have
	\begin{equation}
	\small 0<\sum_{i\neq k}\mathrm{p}_\mathrm{d}(L_i|\epsilon_k^*) <<\mathrm{p}_\mathrm{d}(L_k|\epsilon_k^*)<1.
	\end{equation}
	Then, we can approximate the linear bounds of ROME by $\overline{\mathbf{p}}_{\mathrm{E}}=\overline{\mathbf{p}}_k$ and $\underline{\mathbf{p}}_{\mathrm{E}}=\underline{\mathbf{p}}_k$.
	Thus, the distortion bound and robustness of ROME can also be approximated as
	\begin{IEEEeqnarray}{rCl}
		\mathbf{B}_{\mathrm{E}} & = & \mathbf{B}_k, \IEEEyesnumber\IEEEyessubnumber*\\ r_{\mathrm{E}} & = & r_k.
	\end{IEEEeqnarray}
	This result suggests that the robustness of ROME is close to that of the base classifier $\mathcal{G}_k$, which outperforms all other base classifiers.
	\item \textbf{Confused Case}.
	This is the case when $\epsilon$ falls on the decision boundary of the MPD, i.e., $\epsilon=\epsilon_k$.
	According to Assumption \ref{mpdAssume}, we have
	\begin{equation}
		\small 0<\sum_{i\neq k,k+1}\mathrm{p}_\mathrm{d}(L_i|\epsilon_k)<<\mathrm{p}_\mathrm{d}(L_k|\epsilon_k)=\mathrm{p}_\mathrm{d}(L_{k+1}|\epsilon_k)<1.
	\end{equation}
	Then, we can approximate the linear bounds of ROME by $\overline{\mathbf{p}}_{\mathrm{E}}=(\overline{\mathbf{p}}_k+\overline{\mathbf{p}}_{k+1})/2$ and $\underline{\mathbf{p}}_{\mathrm{E}}=(\underline{\mathbf{p}}_k+\underline{\mathbf{p}}_{k+1})/2$.
	Next, we can compute the output distortion bound of ROME w.r.t. all the perturbed input $\tilde{\mathbf{F}}$ and derive
	\begin{IEEEeqnarray}{rCl}
		\mathbf{B}_{\mathrm{E}} & \preceq &  (\mathbf{B}_k+\mathbf{B}_{k+1})/2, \IEEEyesnumber\IEEEyessubnumber*\\
		r_{\mathrm{E}} & \geq & (r_k+r_{k+1})/2.
	\end{IEEEeqnarray}
	This result indicates that ROME handles attacks with confusing power levels by adjusting its robustness to a median value between $r_k$ and $r_{k+1}$.
	This adjustment effectively mitigates the impact of MPD's accuracy degradation on decision boundaries.
	\item \textbf{General Case}.
	This is the case when $\epsilon$ falls on neither the MPD's decision boundary nor the center of a jamming power level.
	According to Assumption \ref{mpdAssume}, we have
	\begin{equation}
		\small 0<\sum_{i\neq k,k+1}\mathrm{p}_\mathrm{d}(L_i|\epsilon)<<\mathrm{p}_\mathrm{d}(L_k|\epsilon)\approx\mathrm{p}_\mathrm{d}(L_{k+1}|\epsilon)<1.
	\end{equation}
	Then, we can approximate the linear bounds of ROME by $\overline{\mathbf{p}}_{\mathrm{E}}=\mathrm{p}_\mathrm{d}(L_{k}|\epsilon)\overline{\mathbf{p}}_k+\mathrm{p}_\mathrm{d}(L_{k+1}|\epsilon)\overline{\mathbf{p}}_{k+1}$ and $\underline{\mathbf{p}}_{\mathrm{E}}=\mathrm{p}_\mathrm{d}(L_{k}|\epsilon)\underline{\mathbf{p}}_k+\mathrm{p}_\mathrm{d}(L_{k+1}|\epsilon)\underline{\mathbf{p}}_{k+1}$.
	Akin to case (ii), we can derive that
	\begin{IEEEeqnarray}{rCl}
		\mathbf{B}_{\mathrm{E}} & \preceq &  \mathrm{p}_\mathrm{d}(L_{k}|\epsilon)\mathbf{B}_k+\mathrm{p}_\mathrm{d}(L_{k+1}|\epsilon)\mathbf{B}_{k+1}, \IEEEyesnumber\IEEEyessubnumber*\\
		r_{\mathrm{E}} & \geq &  \mathrm{p}_\mathrm{d}(L_{k}|\epsilon)r_k+\mathrm{p}_\mathrm{d}(L_{k+1}|\epsilon)r_{k+1}.
	\end{IEEEeqnarray}
\end{itemize}
	This result demonstrates that ROME can smoothly adjust its robustness across the entire robustness spectrum covered by all the base classifiers.

The analysis above reveals that the proposed ROME framework adaptively adjusts its robustness in response to attacks of varying jamming power. 
Moreover, despite the discrete nature of the pre-defined jamming power levels and the robustness degrees of the base classifiers, ROME manages to achieve a continuous and smooth adjustment of its robustness, delivering sound and promising performance.

\renewcommand{\arraystretch}{1.2}
\begin{table}[htbp]\small
	\centering
	\caption{Settings of the employed networks.}  
	\label{Setting}
	\begin{tabular}{|c|c|c|}
		\hline
		& Layer Name & Dimension  \\
		\hline
		\multirow{5}{*}{\thead{Semantic \\ Encoder \\ \& \\ APG}}
		& ConvLayer & $64$ (kernels) \\
		\cline{2-3}
		& ResBlock & $64$ (kernels) \\
		\cline{2-3}
		& $2\times$ ResBlock & $128$ (kernels) \\
		\cline{2-3}
		& $2\times$ ResBlock & $256$ (kernels) \\
		\cline{2-3}
		& ResBlock & 24 (kernels)  \\
		
		\hline
		\multirow{3}*{\thead{Base \\ Classifiers}}
		& $2\times$ ResBlock & $256$ (kernels) \\
		\cline{2-3}
		& AvgPool & $4$ \\
		\cline{2-3}
		& Dense & $10$ \\
		
		\hline
		\multirow{4}*{\thead{MPD}}
		& LPPG & $10$ \\
		\cline{2-3}
		& $4\times$ ResBlock & $256$ (kernels) \\
		\cline{2-3}
		& AvgPool & $2$ \\
		\cline{2-3}
		& Dense & $4$  \\
		\hline
	\end{tabular}
\end{table}

%
%

\section{Simulation Results} \label{Simulation}

\begin{figure}[htbp]
	\centering
	\includegraphics[width=0.4\textwidth]{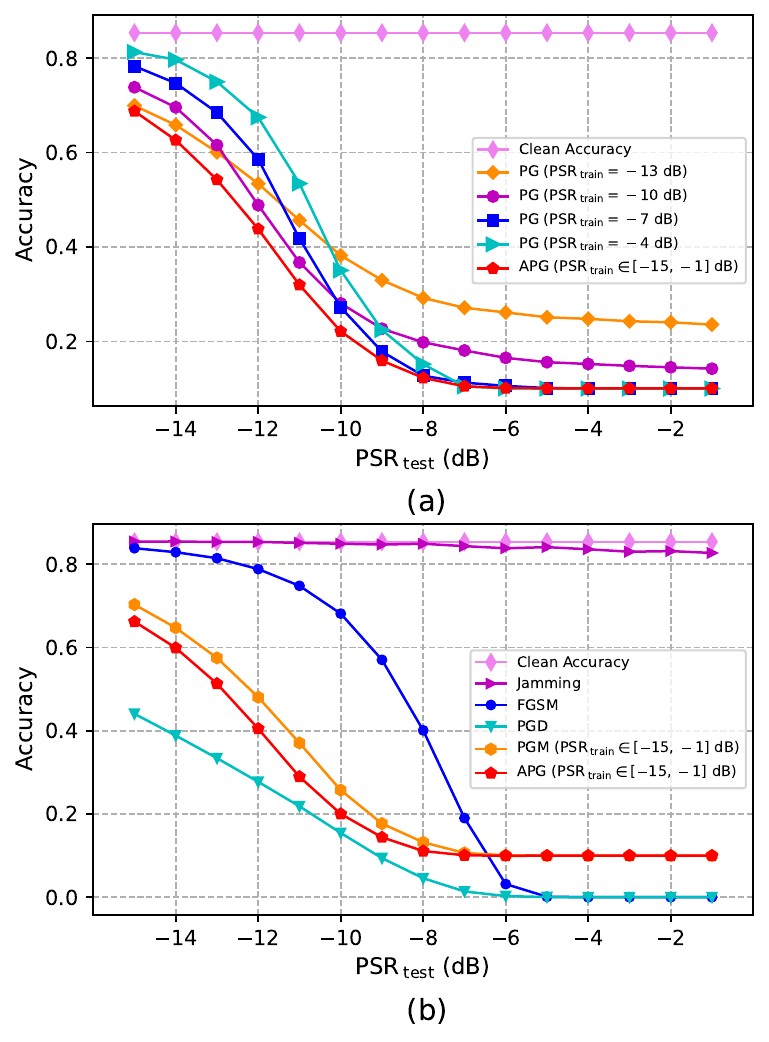}
	\captionsetup{font=footnotesize}
	\caption{Performance of the APG and comparison with existing attack models.}
	\label{APGeffect}
\end{figure}

%
%
%
%

\begin{figure}[htbp]
	\centering
	\includegraphics[width=0.4\textwidth]{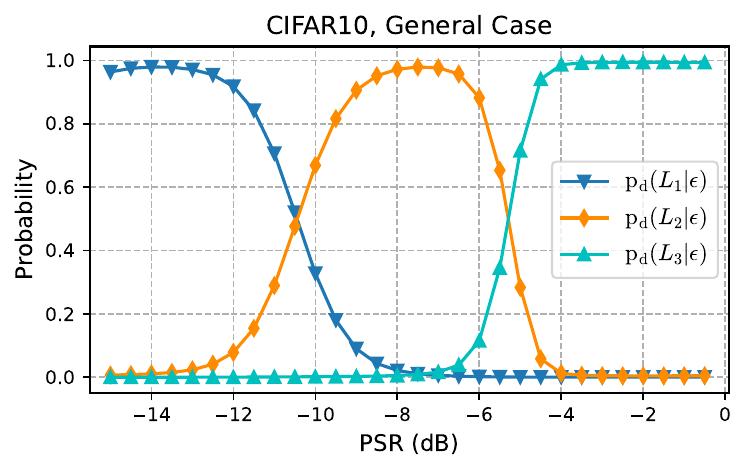}
	\captionsetup{font=footnotesize}
	\caption{Statistics of the MPD's output. The experiment is conducted in the general case setting on the CIFAR10 dataset.}
	\label{PdSim}
\end{figure}

\begin{figure*}[t]
	\centering
	\includegraphics[width=0.9\textwidth]{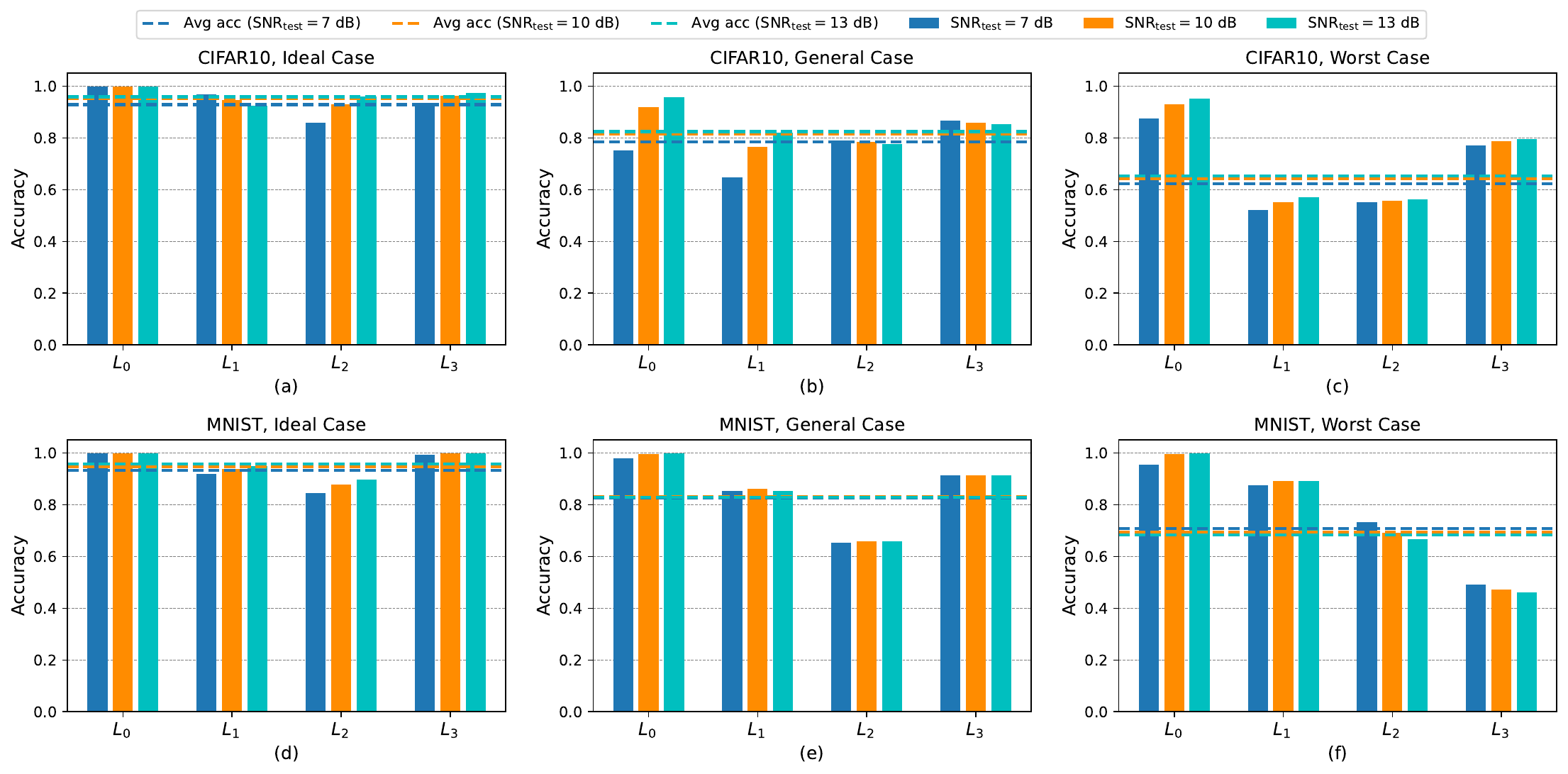}
	\captionsetup{font=footnotesize}
	\caption{Detection performance of the proposed MPD. The MPD is evaluated across different cases and power levels on the CIFAR10 and MNIST datasets at three different $\mathrm{SNR}$s: $7$ dB, $10$ dB, $13$ dB. We provide the average detection accuracy scores on each power level label across $10,000$ samples in each test dataset, along with their mean values.}
	\label{mpdcases}
\end{figure*}

\begin{figure}[t]
	\centering
	\includegraphics[width=0.43\textwidth]{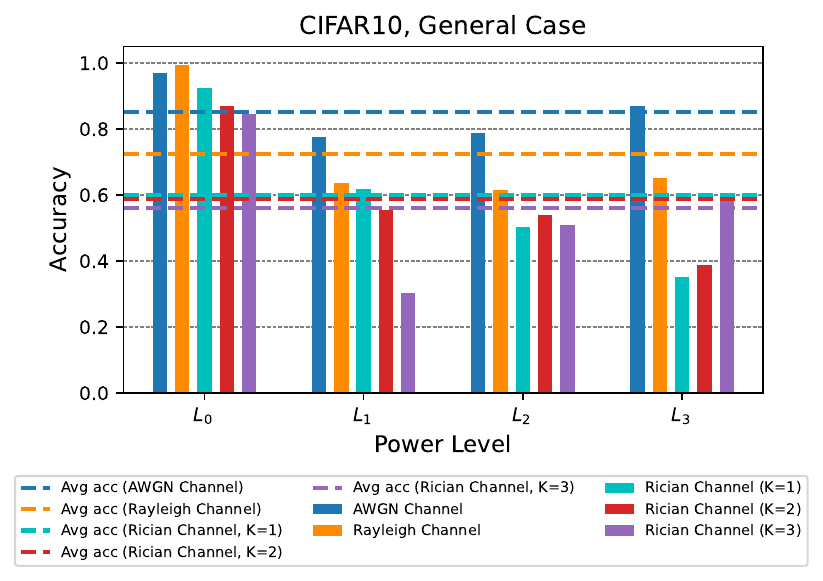}
	\captionsetup{font=footnotesize}
	\caption{Performance of the MPD across various channel models. The evaluation is conducted in the general case setting on the CIFAR10 dataset.}
	\label{mpdfading}
\end{figure}

\begin{figure*}[t]
	\centering
	\includegraphics[width=0.9\textwidth]{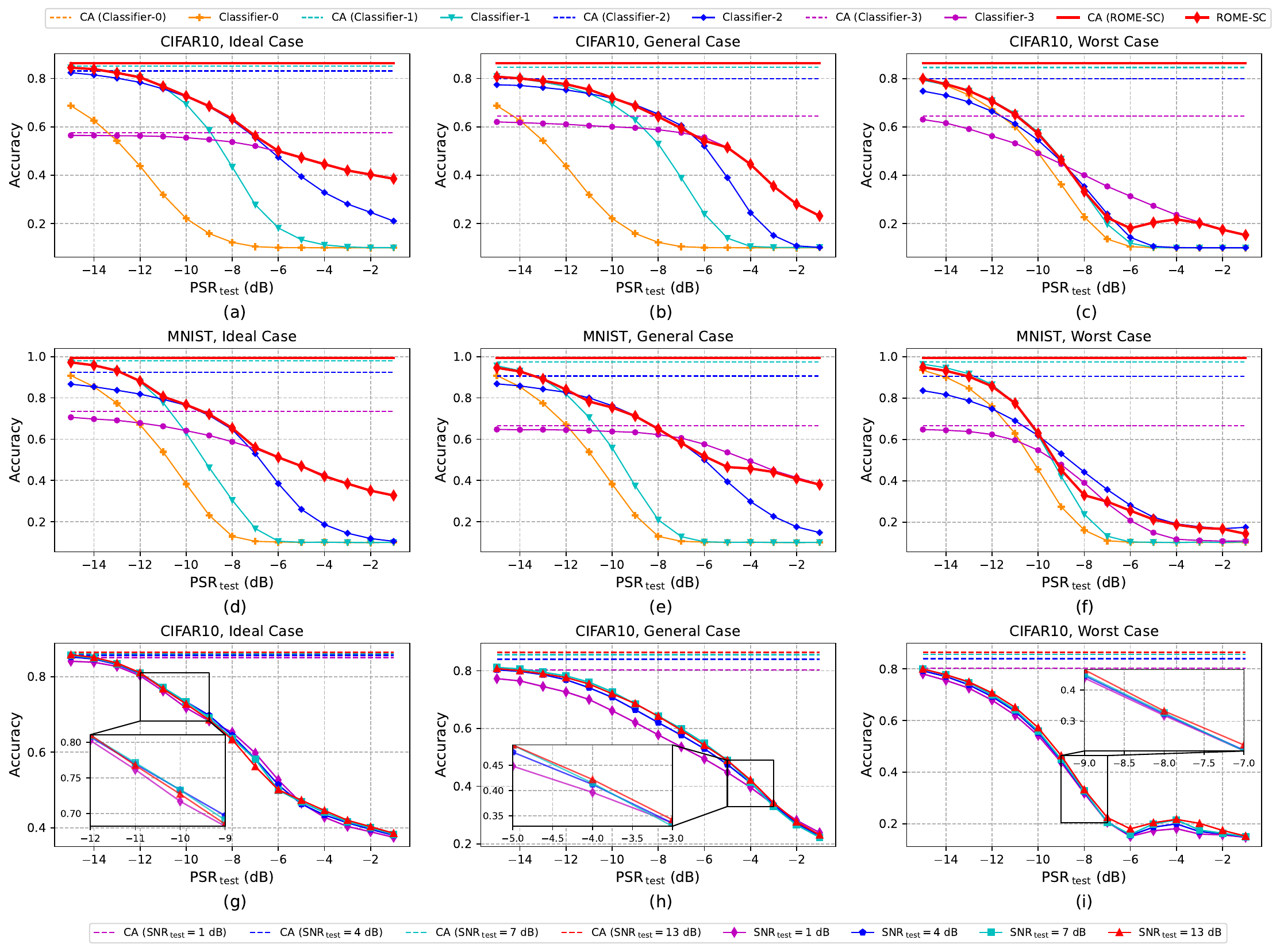}
	\captionsetup{font=footnotesize}
	\caption{Performance of the proposed ROME framework. (a)-(f) present the case study results by examining the performance of ROME and each base classifier across different cases on the CIFAR10 and MNIST datasets. (g)-(i) assess the impact of $\mathrm{SNR}$ variations by examining ROME across different cases on the CIFAR10 dataset at four different $\mathrm{SNR}$s: $1$ dB, $4$ dB, $7$ dB, $13$ dB. We provide the average classification accuracy scores across $10,000$ images in each test dataset at various $\mathrm{PSR}$s within a range of $[-15, -1]$ dB.}
	\label{syscases}
\end{figure*}

\subsection{Settings}
In this section, we examine the performance of our proposed APG, MPD, and ROME.
The simulations are conducted using the PyTorch platform.
We construct the DNN models based on convolutional neural networks (CNNs).
To train the basic models, we adopt the ``Adam'' optimizer with an initial learning rate set to $0.0001$.
The learning rate is gradually reduced by the cosine annealing algorithm \cite{Loshchilov2017sgdr} as training progresses.
The batch size is set to $128$.
{The system is evaluated on the MNIST and CIFAR-10 datasets, two public datasets that are widely adopted in the realm of image processing.
The MNIST dataset comprises grayscale images of handwritten digits across $10$ classes, with $60, 000$ images of size $28\times28$ in the training set and $10, 000$ in the test set.
The CIFAR10 dataset consists of RGB images of size $32\times 32$ across $10$ classes, with $50, 000$ images in the training dataset and $10, 000$ in the testing dataset.
Furthermore, the system evaluations are conducted over various channel models, including AWGN, Rayleigh, and Rician channels.}
We set $\mathrm{SNR}=$ $13$ dB during the training stage.
The jamming power is assumed to vary within $\mathrm{PSR}\in$ $[-15,-1]$ dB, where $\mathrm{PSR}$ refers to perturbation-to-signal ratio, defined as
\begin{IEEEeqnarray}{rCl}
	\mathrm{PSR} = 10 \log _{10}\frac{\epsilon^2}{P}.
\end{IEEEeqnarray}
Besides, we set $N=4$ base classifiers at the receiver.
Correspondingly, the jamming power is divided into four levels (i.e., $V=\{L_0,L_1,L_2,L_3\}$) based on $\mathrm{PSR}$.
{The settings of the employed networks are presented in Table \ref{Setting}.
Below, we provide detailed settings for the case study, categorized from the perspective of the legitimate user.
\begin{itemize}
	\item \textbf{Ideal Case:}
	The attacker's APG model is trained using the base classifier $\mathcal{G}_0$, and all models at the legitimate receiver are trained to counter the APG.
	The power level definition is tailored for the base classifiers to counter the APG.
	\item \textbf{General Case:}
	The attacker's APG model is trained using the base classifier $\mathcal{G}_0$, while all models at the legitimate receiver are trained to counter an agent attack model, PGD.
	The power level definitions align with those used during the training of the base classifiers.
	\item \textbf{Worst Case:}
	The attacker's APG model is trained against all models at the legitimate receiver, which are trained to counter an agent attack model, PGD.
	The power level definition is consistent with the power levels utilized in the training of the base classifiers.
\end{itemize}}

{\subsection{Attack Performance of APG}}
Fig. \ref{APGeffect} (a) illustrates the effectiveness of APG in generating power-variable attacks.
The target model is the standard classifier $\mathcal{G}_0$ without AT defense, with the upper bound representing its average classification accuracy on clean samples.
``PG'' denotes a perturbation generator that operates without the PAM and is trained with a fixed $\mathrm{PSR}$.
It is evident that our APG delivers the most destructive attacks across the entire $\mathrm{PSR}_{\mathrm{test}}$ region.
Notably, when $\mathrm{PSR}_{\mathrm{test}}$ is $-7$ dB, the accuracy drops to $10.00\%$, a sign that the classifier is essentially making random guesses.
These results stem from the design of PAM, which equips our APG with foresight about the upcoming power constraint throughout the perturbation generation process.
Besides, we observe that a PG's performance declines when $\mathrm{PSR}_{\mathrm{test}}$ diverges from $\mathrm{PSR}_{\mathrm{train}}$, especially when the discrepancy between $\mathrm{PSR}_{\mathrm{test}}$ and $\mathrm{PSR}_{\mathrm{train}}$ is significant, which aligns with expectations.

{Additionally, we compare our APG with other attack methodologies.
All the methods are evaluated on the CIFAR10 dataset by targeting the standard classifier $\mathcal{G}_0$. 
\begin{itemize}
	\item \textit{Jamming:} A fundamental jamming attack that adheres to the Gaussian distribution. It is semantic-irrelevant and widely utilized in conventional jamming scenarios.
	\item \textit{FGSM} \cite{Goodfellow2014explaining}: A standard gradient-based technique for crafting adversarial examples. We employ the publicly available FGSM implementation in our experiments.
	\item \textit{PGD} \cite{Madry2018towards}: Another popular gradient-based attack method, distinct from the FGSM approach. We utilize the open-source PGD implementation to generate this attack.
	\item \textit{PGM} \cite{Bahramali2021robust}: An attack model designed for generating input-agnostic semantic jamming signals. We re-implement this baseline as it is not publicly available. To ensure fairness, the PGM is trained with variable power constraint within $\mathrm{PSR}_{\mathrm{test}}\in[-15,-1]$ dB as APG does.
\end{itemize}
As shown in Fig. \ref{APGeffect} (b), the conventional jamming attack exerts minimal impact on the legitimate user's system performance.
This is because the jamming signal is not generated with the aim of semantic distortion, thus posing limited threats to SC systems.
Moreover, a gap in attack effectiveness is observed between the APG and PGD attacks.
This result is rationalized by the fact that APG satisfies the input-agnostic constraint in wireless attack scenarios, which forces semantic jamming attacks to operate as gray-box attacks.
Conversely, the PGD is a kind of white-box attacks that neglects the input-agnostic constraint.
Despite its good performance, the PGD method is impractical for generating semantic jamming signals and is only suitable for offline simulation analysis.
Moreover, it is observed that the proposed APG attack outperforms the PGM method across the entire $\mathrm{PSR}_{\mathrm{test}}$ region.
This further underscores the superiority of APG in generating effective semantic jamming attacks under diverse jamming power requirements.}\\

{
\subsection{Detection Performance of MPD}
Fig. \ref{PdSim} shows the statistics of the MPD's output.}
Each data point represents the average result computed from $10,000$ samples with a specific PSR.
Compared to the analysis in Section \ref{Analysis}, these results demonstrate that the MPD meets the requirements for supporting the operation of ROME, which is consistent with our expectations.

\begin{figure}[htbp]
	\centering
	\includegraphics[width=0.4\textwidth]{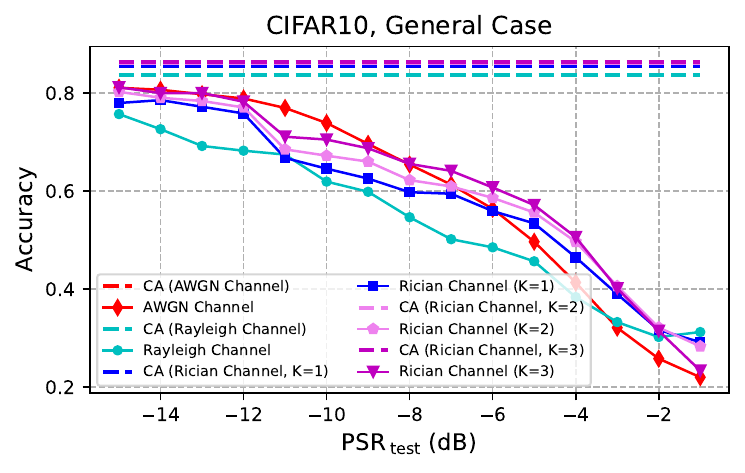}
	\captionsetup{font=footnotesize}
	\caption{System performance across various channel types. }
	\label{sysfading}
\end{figure}

{Fig. \ref{mpdcases} presents the evaluation results of the MPD's detection performance.
We observe that the $\mathrm{SNR}$ mismatch between training and deployment stages holds negligible impact on the MPD's performance.
This observation implies that the MPD's capability of identifying jamming power levels remains largely unaffected by channel noises, underscoring the MPD's robustness and reliability.
This robustness is attributed to our design of the LPPG module.
While the perturbations in the raw received signal can be a mixture of the jamming signal and channel noises, complicating direct identification of jamming power from the raw signal, the LPPG module offers an auxiliary stream of information by monitoring the behavior of the base classifiers.
This enables the MPD to more effectively identify jamming signals amidst channel noise interference, enhancing its performance across varying $\mathrm{SNR}$ conditions.
As observed in Fig. \ref{mpdcases} (a), (b), (d), and (e), the MPD trained with an agent attack model, PGD, maintains comparable detection accuracy to the ideal case where it is trained with the exact attack model.
This indicates the MPD's reliability in detecting the presence and power level of unknown attacks.
Even in the worst cases shown in Fig. \ref{mpdcases} (c) and (f), the MPD achieves an average accuracy of $62.22\%$ at $\mathrm{SNR}=7$ dB on CIFAR10, and $70.91\%$ on MNIST, which further demonstrates the robustness of the proposed MPD.

Fig. \ref{mpdfading} illustrates the detection performance of the MPD across various channel models.
Specifically, we compare the MPD's performance in AWGN, Rayleigh, and Rician channels where the Rican factors for Rician channels are set to $K=1,2$, and $3$.
It is observed that the MPD's detection accuracy diminishes in fading channels compared to the AWGN channel.
This degradation is expected, as fading effects can significantly affect semantic jamming signals, causing them to deviate from their initial power levels and lose distinct jamming-related patterns.
Nevertheless, the fading effects also contribute to the legitimate user's advantage, as the impact of semantic jamming signals is similarly attenuated by channel fading.
This insight is supported by the results provided in Fig. \ref{sysfading}, which indicate that the overall performance of ROME remains remarkable in fading channels, despite the degraded detection accuracy of the MPD.}

{\subsection{Evaluation of ROME}
Fig. \ref{syscases} (a)-(f) illustrates the performance of ROME across different cases and $\mathrm{PSR}$s on the CIFAR10 and MNIST datasets.}
It is apparent that the ROME approach achieves superior accuracy, whether on clean samples (CA) or across the entire $\mathrm{PSR}_{\mathrm{test}}$ region.
This success is attributed to the system's capability of balancing robustness with generalization ability.
Notably, even when the $\mathrm{PSR}$ approaches the MPD's decision boundaries, our ROME approach maintains remarkable performance.
This resilience is attributed to our model ensembling approach, {which leverages the soft decisions (the confidence on each power level) from the MPD, rather than relying exclusively on hard decisions (a single, definitive power level).}
While the hard decision accuracy suffers a decline at decision boundaries, the soft decision can accurately reflect that the neighboring power levels share similar probabilities.
Consequently, the accuracy decrease at decision boundaries for hard decisions has only negligible effects on the overall system performance.
{Specifically, Fig. \ref{syscases} (c) and (f) present the performance of ROME in the worst case.
Even in such scenarios, ROME demonstrates remarkable performance.
We observe that the performance of the base classifiers converges under worst-case attacks.}
This convergence occurs because the improved APG learns to equivalently mislead the base classifiers, so as to reduce the overall performance of RMOE-SC to its worst value.
{Additionally, in Fig. \ref{syscases} (c) there is a notable degradation in the system performance curve when $\mathrm{PSR}_{\mathrm{test}}=[-9,-4]$ dB, primarily due to the non-ideal jamming power level definition.}
Despite all these adverse factors, ROME maintains a strong capability in balancing its robustness and generalization ability, which further demonstrates its superiority in handling semantic jamming attacks.

{Fig. \ref{syscases} (g)-(i) investigate the impact of $\mathrm{SNR}$ variations on the performance of ROME across different cases and $\mathrm{PSR}$s on the CIFAR10 dataset.
It is evident that $\mathrm{SNR}$ variations have negligible impacts on the performance of ROME in the ideal case.
In terms of the general and worst cases, the influence of hash channel conditions is still very limited.}
Even in scenarios with considerable $\mathrm{SNR}$ mismatch, such as $\mathrm{SNR}_{\mathrm{test}}=$ $1$ dB, ROME's accuracy reduction is no more than $8.10\%$ across the entire $\mathrm{PSR}$ range.
These observations demonstrate that ROME also possesses remarkable robustness against channel noise, and can sustain its superiority in harsh channel conditions.

\begin{figure}[htbp]
	\centering
	\includegraphics[width=0.4\textwidth]{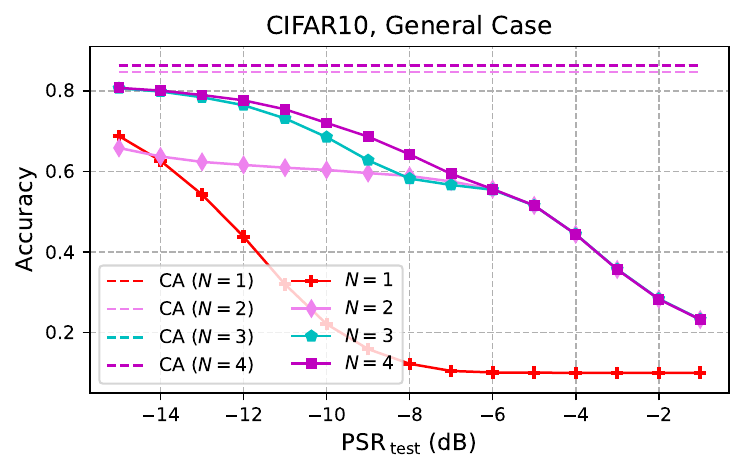}
	\captionsetup{font=footnotesize}
	\caption{System performance with different numbers of base classifiers. }
	\label{BaseNum}
\end{figure}

\begin{figure}[t]
	\centering
	\includegraphics[width=0.4\textwidth]{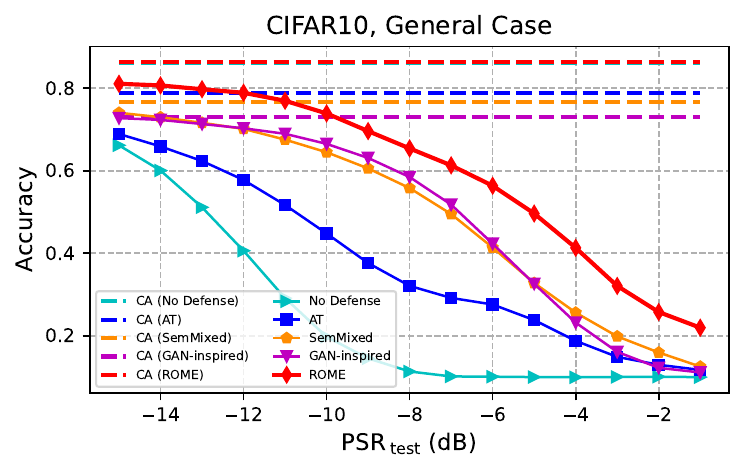}
	\captionsetup{font=footnotesize}
	\caption{System performance across various channel types. }
	\label{syscompare}
\end{figure}

Fig. \ref{BaseNum} shows the impact of the number of base classifiers on the performance of ROME.
We consider the general case { on the CIFAR10 dataset} and observe that the performance improvements converge quickly as the number of base classifiers increases.
Intuitively, increasing the number of base classifiers could enhance the system performance, while the observation indicates that it is unnecessary to employ too many base classifiers due to diminishing marginal benefits.
Generally, setting the number of base classifiers $N$ to $4$ is adequate for ROME to achieve excellent performance.
Consequently, implementing ROME does not impose significant storage overhead.

{Fig. \ref{sysfading} provides the performance of ROME across various channel types.
Specifically, we conduct evaluations in AWGN, Rayleigh, and Rician channels, with Rician factors set at $K=1,2$, and $3$.
The results suggest that the situations in fading channels are not fundamentally different from that in the AWGN channel.
Specifically, it is observed that an increase in the $K$ factor improves ROME performance, despite the MPD's diminished detection precision observed in Fig. \ref{mpdfading}.
Moreover, under high-power semantic jamming attacks, ROME exhibits better performance in fading channels compared to the AWGN channel.
These findings are attributed to the fact that fading impacts both the efficacy of semantic jamming attacks and the precision of jamming detection.
Thus, the strategic balance established between the attacker and the user in the AWGN channel is preserved even when fading effects are taken into account.

Fig. \ref{syscompare} compares ROME's performance with existing defense methodologies on the CIFAR10 dataset, using APG as the attack model trained on the standard classifier $\mathcal{G}_0$.
To ensure fairness, we train the baselines with semantic jamming signals across a range of power levels, specifically within the $\mathrm{PSR}_{\mathrm{train}}$ range of $[-15, -1]$ dB, mirroring the training of ROME's base classifiers.
The baselines are detailed as follows.
\begin{itemize}
	\item \textit{No Defense} : Use $\mathcal{G}_0$ without AT.
	\item \textit{AT} : Perform AT to $\mathcal{G}_0$ with only the PGD attacks.
	\item \textit{SemMixed} \cite{Nan2023physical}: Perform AT to $\mathcal{G}_0$ with multiple types of adversarial attacks and clean samples. We employ the PGM and PGD attacks, which are randomly selected during the training.
	\item \textit{GAN-Inspired Strategy} \cite{Tang2023gan}: Train $\mathcal{G}_0$ with an agent PG and a two-class jamming detector using GAN-like training strategy.
\end{itemize}
The results indicate that all the baselines only achieves moderate performance when encountering power-variable semantic jamming attack, despite considering these in their AT process.
This limitation stems from their offline training approach, which lacks real-time adaptive mechanisms to handle dynamic attacks.
In contrast, ROME consistently demonstrates graceful performance both on clean samples and under power-variable attacks, showcasing its superiority and reliability.}

{\subsection{Complexity Analysis}
We now provide a complexity analysis of our methods, encompassing the computational complexity, storage requirements, and practical runtime for all the proposed algorithms.
%
\begin{itemize}
	\item \textbf{\textit{APG:}}
	The APG's most computationally intensive operations are 2D convolutions and weighted summations in the FC layers.
	The complexity of a single convolutional layer is $\mathcal{O}(K^2C_iC_oH_oW_o)$ \cite{Howard2017mobilenets}, where $K$, $C_i$, and $C_o$ are constants, and $H_oW_o$ is dependent on the input image size $H_{in}W_{in}$.
	Therefore, the complexity of 2D convolutions scales with $\mathcal{O}(H_{in}W_{in})$.
	Additionally, the complexity of a single FC layer is $\mathcal{O}(D_iD_o)$ \cite{Sze2017efficient}, and since $D_i$ and $D_o$ are both constants, the total complexity of the FC layers is $\mathcal{O}(1)$.
	In summary, the overall complexity of the APG is $\mathcal{O}(H_{in}W_{in})$.
	\item \textbf{\textit{Semantic Encoder:}}
	The semantic encoder is composed entirely of convolutional layers, without any FC layers.
	As a result, its primary operations are 2D convolutions, yielding an overall complexity of $\mathcal{O}(H_{in}W_{in})$.
	\item \textbf{\textit{Base Classifiers:}}
	In the base classifiers, the main operations are 2D convolutions and weighted summations in the FC layers.
	Given an input size $H_{out}W_{out}$, which depends on $H_{in}W_{in}$, the complexity of the convolutional layers is $\mathcal{O}(H_{in}W_{in})$.
	Since $D_i$ and $D_o$ are constants, the overall complexity of the FC layers is $\mathcal{O}(1)$.
	As the base classifiers execute computations in parallel, the total complexity of all the base classifiers is $\mathcal{O}(H_{in}W_{in})$.
	\item \textbf{\textit{MPD:}}
	The computational cost in the MPD primarily arises from 2D convolutions and weighted summations in the FC layers.
	Consequently, the complexity analysis is similar to that of the base classifiers, leading to a total complexity of $\mathcal{O}(H_{in}W_{in})$.
\end{itemize}
The algorithm complexity of ROME primarily comes from three components: the semantic encoder, the base classifiers, and the MPD.
In conclusion, the overall algorithm complexity of ROME is $\mathcal{O}(H_{in}W_{in})$.\\

\renewcommand{\arraystretch}{1.2}
\begin{table}[t]\scriptsize
	\centering
	\caption{Model size and inference time on CIFAR-10}  
	\label{complexity2}
	\begin{tabular}{l|c|c}
		\toprule
		\textbf{Model} & \textbf{Size (Params)} & \textbf{Inference Time (Per Image)} \\
		\midrule
		\midrule
		APG & $3.020$ M & $3.585\times10^{-5}$ s \\
		Semantic Encoder & $2.658$ M & $1.229\times 10^{-5}$ s  \\
		Base Classifier & $1.836$ M & $4.173\times 10^{-6}$ s \\
		MPD & $3.748$ M & $1.332\times 10^{-5}$ s \\
		ROME (4 Base Classifiers) & $13.75$ M & $3.263\times 10^{-5}$ s \\
		JSCC Encoder \cite{Yang2022ofdm} & $2.791$ M & $1.440\times 10^{-5}$ s \\
		JSCC Decoder \cite{Yang2022ofdm} & $2.790$ M & $1.159\times 10^{-5}$ s \\
		JSCC \cite{Yang2022ofdm} & $5.581$ M & $2.598\times 10^{-5}$ s \\
		\bottomrule
	\end{tabular}
\end{table}

To conclude our discussion on computational complexity and practical storage considerations, we have detailed the number of model parameters and  measured the average inference time of ROME on a Linux server with two $2.60$ GHz Intel(R) Xeon(R) Platinum 8350C CPUs and four NVIDIA GeForce RTX 4090 GPUs.
Each experiment was conducted using $64$ CPU threads and a single GPU on the CIFAR-10 dataset.
The results are summarized in Table \ref{complexity2}.
For comparison, we also measured the average inference time of a prior work, JSCC \cite{Yang2022ofdm}, using the same GPU implementation as ROME.
These results demonstrate that ROME's computational and storage demands are in line with those of existing deep learning-based methods, which have been proven to be significantly more efficient than traditional separate source and channel coding approaches \cite{Bourtsoulatze2019deep}.
Thus, the ROME framework holds the practicality for real-world deployment.
}
\section{Conclusion} \label{Conclusion}
In this paper, we introduced a novel framework for robust model ensembling in SCs.
Our approach aims to dynamically balance the trade-off between robustness and generalization ability, thereby enhancing the overall system performance.
First, we presented the APG, a semantic jamming attack method capable of generating adversarial perturbations spanning a broad spectrum of power levels while maintaining destructive characteristics.
Following this, we introduced the MPD to detect the presence of semantic jamming attacks and measure their power levels.
With the MPD in place, we designed a robust model ensembling approach guided by the MPD.
Subsequently, we conducted a theoretical analysis of the distortion bound of ROME, formulated its robustness, and demonstrated its capacity for adaptive robustness.
Simulation results exhibit that the proposed method significantly enhances the overall performance of SC systems in the presence of semantic jamming attacks.

\section*{Appendix A\\Derivation of Maximum Upper Bound}
The derivation utilizes the properties of the dual norm, whose definition is given as follows.
\begin{definition}\label{dualnorm}
	\textbf{(Dual Norm)}
	
	Let $||\cdot||$ be a norm on $\mathbb{R}^n$, $\mathbf {x}=[\mathrm{x}_1,...,\mathrm{x}_n]^T\in \mathbb{R}^n$, and $\mathbf {z}=[\mathrm{z}_1,...,\mathrm{z}_n]^T\in \mathbb{R}^n$.
	The corresponding dual norm, denoted by $||\cdot||_*$, is defined as
	\begin{equation}
		||\mathbf {z}||_* = \sup\{\mathbf {z}^T\mathbf {x}: ||\mathbf {x}||\leq 1\}.
	\end{equation}
\end{definition}
Then, we present the detailed derivation of the maximum upper bound.
The minimum lower bound can be derived similarly.
{\small\begin{IEEEeqnarray}{rCl}
	\max_{\tilde{\mathbf{F}}}\overline{\mathbf{p}} & = & \max_{\tilde{\mathbf{F}}}\overline{\mathbf{W}}_n\tilde{\mathbf{F}}+\overline{\mathbf{b}}_n \IEEEnonumber \\
	& = & (\epsilon+\eta)\max_{\mathbf{J}\in \mathbb{B}_p(\mathbf{0}, 1)}\overline{\mathbf{W}}_n\mathbf{J}+\overline{\mathbf{W}}_n\mathbf{F}+\overline{\mathbf{b}}_n \IEEEyesnumber \IEEEyessubnumber \label{maxupper1}\\
	& = & (\epsilon+\eta)||\overline{\mathbf{W}}_n||_q+\overline{\mathbf{W}}_n\mathbf{F}+\overline{\mathbf{b}}_n \IEEEyessubnumber.  \label{maxupper2}
\end{IEEEeqnarray}}
The derivation from Eq. (\ref{maxupper1}) to Eq. (\ref{maxupper2}) utilizes the properties of dual norm.
According to Definition \ref{dualnorm}, the term $\max_{\mathbf{J}\in \mathbb{B}_p(\mathbf{0}, 1)}\overline{\mathbf{W}}_n\mathbf{J}$ is equivalent to $||\overline{\mathbf{W}}_n||_*$, where the $||\cdot||_*$ here denotes the dual norm of the $\ell_p$-norm on $\mathbb{R}^n$.

We next present the proof that the dual norm of the $\ell_p$-norm is the $\ell_q$-norm, where $\frac{1}{p}+\frac{1}{q}=1$.
According to Definition \ref{dualnorm}, the dual norm of the $\ell_p$-norm is defined as
\begin{equation}
	||\mathbf{z}||_*=\sup\{\mathbf{z}^T\mathbf{x}: ||\mathbf{x}||_p\leq 1\}.
\end{equation}
By Hölder's inequality, we have
\begin{equation}
	\mathbf {z}^T\mathbf{x}\leq||\mathbf{z}||_q||\mathbf{x}||_p.
\end{equation}
Since $||\mathbf {x}||_p\leq 1$, we have $||\mathbf{z}||_*\leq||\mathbf{z}||_q$.

Then, to prove $||\mathbf{z}||_*=||\mathbf {z}||_q$, we only need to prove $||\mathbf{z}||_*\geq||\mathbf {z}||_q$.
To this end, we choose $\tilde{\mathbf{x}}=\frac{|\mathbf{z}|^{q-1}\mathrm{sgn}(\mathbf{z})}{||\mathbf{z}||_q^{q-1}}$, then we have
{\small\begin{IEEEeqnarray}{rCl}
	||\tilde{\mathbf{x}}||_p & = & (\sum^n_{i=1}|\frac{|\mathrm{z}_i|^{q-1}\mathrm{sgn}(\mathrm{z}_i)}{||\mathbf{z}||_q^{q-1}}|^p)^{\frac{1}{p}} \IEEEnonumber \\
	& = & \frac{1}{||\mathbf{z}||_q^{q-1}}(\sum^n_{i=1}|\mathrm{z}_i|^q)^{\frac{1}{p}} \IEEEnonumber \\
	& = & \frac{1}{||\mathbf{z}||_q^{q-1}}((\sum^n_{i=1}|\mathrm{z}_i|^q)^{\frac{1}{q}})^{\frac{q}{p}} \IEEEnonumber\\
	& = & \frac{||\mathbf{z}||_q^{q-1}}{||\mathbf{z}||_q^{q-1}}=1,
\end{IEEEeqnarray}}
which satisfies $||\mathbf{x}||_p\leq 1$.
Therefore, we have
\begin{equation}
	||\mathbf{z}||_* \geq \mathbf{z}^T\tilde{\mathbf{x}}=\frac{\sum_{i=1}^n|\mathrm{z_i}|^q}{||\mathbf{z}||_q^{q-1}}=||\mathbf{z}||_q.
\end{equation}
\section*{Appendix B\\Derivation of ROME's Bound and Robustness}
We present the detailed derivation of ROME's bound and robustness considering the \textbf{General Case}.
The \textbf{Confident case} and \textbf{Confused case} are special cases of the \textbf{General Case} and the conclusions can be derived in a similar manner.

We first compute the maximum upper bound of ROME w.r.t. all the perturbed input $\tilde{\mathbf{F}}$, given as
{\small\begin{IEEEeqnarray}{rCl}
	\max_{\tilde{\mathbf{F}}}\overline{\mathbf{p}}_E & = & \max_{\tilde{\mathbf{F}}}(\mathrm{p}_\mathrm{d}(L_k|\epsilon)\overline{\mathbf{W}}_n^{(k)}+\mathrm{p}_\mathrm{d}(L_{k+1}|\epsilon)\overline{\mathbf{W}}_n^{(k+1)})\tilde{\mathbf{F}}\IEEEnonumber\\
	& & + (\mathrm{p}_\mathrm{d}(L_k|\epsilon)\overline{\mathbf{b}}_n^{(k)}+\mathrm{p}_\mathrm{d}(L_{k+1}|\epsilon)\overline{\mathbf{b}}_n^{(k+1)}) \IEEEnonumber \\
	& = & (\epsilon+\eta)||\mathrm{p}_\mathrm{d}(L_k|\epsilon)\overline{\mathbf{W}}_n^{(k)}+\mathrm{p}_\mathrm{d}(L_{k+1}|\epsilon)\overline{\mathbf{W}}_n^{(k+1)}||_q\IEEEnonumber\\
	& & + (\mathrm{p}_\mathrm{d}(L_k|\epsilon)\overline{\mathbf{W}}_n^{(k)}+\mathrm{p}_\mathrm{d}(L_{k+1}|\epsilon)\overline{\mathbf{W}}_n^{(k+1)})\mathbf{F}\IEEEnonumber\\
	& & + (\mathrm{p}_\mathrm{d}(L_k|\epsilon)\overline{\mathbf{b}}_n^{(k)}+\mathrm{p}_\mathrm{d}(L_{k+1}|\epsilon)\overline{\mathbf{b}}_n^{(k+1)}) \IEEEyesnumber.
\end{IEEEeqnarray}}\noindent
The minimum lower bound can be derived in a similar way.
{Then, we have
	{\small
\begin{IEEEeqnarray}{rCl}
	\mathbf{B}_{\mathrm{E}} & = & (\epsilon+\eta)(||\sum_{i=k}^{k+1}\mathrm{p}_\mathrm{d}(L_i|\epsilon)\overline{\mathbf{W}}_n^{(i)}||_q + ||\sum_{i=k}^{k+1}\mathrm{p}_\mathrm{d}(L_i|\epsilon)\underline{\mathbf{W}}_n^{(i)}||_q) \IEEEnonumber*\\
	& & + \sqrt{kP}||\sum_{i=k}^{k+1}\{\mathrm{p}_\mathrm{d}(L_i|\epsilon)\overline{\mathbf{W}}_n^{(i)} - \mathrm{p}_\mathrm{d}(L_i|\epsilon)\underline{\mathbf{W}}_n^{(i)}\}||_2  \\
	& & + \sum_{i=k}^{k+1}\{(\mathrm{p}_\mathrm{d}(L_i|\epsilon)\overline{\mathbf{b}}_n^{(i)} - \mathrm{p}_\mathrm{d}(L_i|\epsilon)\underline{\mathbf{b}}_n^{(i)})\}\IEEEyesnumber\IEEEyessubnumber\label{be1}\\
	& \preceq & \sum_{i=k}^{k+1}\{(\epsilon+\eta)(||\mathrm{p}_\mathrm{d}(L_i|\epsilon)\overline{\mathbf{W}}_n^{(i)}||_q + ||\mathrm{p}_\mathrm{d}(L_i|\epsilon)\underline{\mathbf{W}}_n^{(i)}||_q) \IEEEnonumber* \\
	& & + \sqrt{kP}||\mathrm{p}_\mathrm{d}(L_i|\epsilon)\overline{\mathbf{W}}_n^{(i)} - \mathrm{p}_\mathrm{d}(L_i|\epsilon)\underline{\mathbf{W}}_n^{(i)}||_2  \\
	& & + (\mathrm{p}_\mathrm{d}(L_i|\epsilon)\overline{\mathbf{b}}_n^{(i)} - \mathrm{p}_\mathrm{d}(L_i|\epsilon)\underline{\mathbf{b}}_n^{(i)})\} \IEEEyessubnumber\label{be2}\\
	& = &\mathrm{p}_\mathrm{d}(L_k|\epsilon)\mathbf{B}_k+\mathrm{p}_\mathrm{d}(L_{k+1}|\epsilon)\mathbf{B}_{k+1}. \IEEEyessubnumber
\end{IEEEeqnarray}}\noindent
The derivation from Eq. (\ref{be1}) to Eq. (\ref{be2}) utilizes the triangle inequality property.
Likewise, the following steps from Eq. (\ref{r1}) to Eq. (\ref{r3}) also relies on this property.
{\small
\begin{IEEEeqnarray}{rCl}
	r_{\mathrm{E}} & = & -||||\sum_{i=k}^{k+1}\mathrm{p}_\mathrm{d}(L_i|\epsilon)\overline{\mathbf{W}}_n^{(i)}||_q +  ||\sum_{i=k}^{k+1}\mathrm{p}_\mathrm{d}(L_i|\epsilon)\underline{\mathbf{W}}_n^{(i)}||_q||_p \IEEEnonumber* \\
	& & - ||||\sum_{i=k}^{k+1}\{\mathrm{p}_\mathrm{d}(L_i|\epsilon)\overline{\mathbf{W}}_n^{(i)} - \mathrm{p}_\mathrm{d}(L_i|\epsilon)\underline{\mathbf{W}}_n^{(i)}\}||_2||_p \\
	& & - ||\sum_{i=k}^{k+1}\{\mathrm{p}_\mathrm{d}(L_i|\epsilon)\overline{\mathbf{b}}_n^{(i)} - \mathrm{p}_\mathrm{d}(L_i|\epsilon)\underline{\mathbf{b}}_n^{(i)}\}||_p \IEEEyesnumber\IEEEyessubnumber\label{r1}\\
	& \geq & -||\sum_{i=k}^{k+1}\{||\mathrm{p}_\mathrm{d}(L_i|\epsilon)\overline{\mathbf{W}}_n^{(i)}||_q + ||\mathrm{p}_\mathrm{d}(L_i|\epsilon)\underline{\mathbf{W}}_n^{(i)}||_q\}||_p \\
	& & - ||\sum_{i=k}^{k+1}||\mathrm{p}_\mathrm{d}(L_i|\epsilon)\overline{\mathbf{W}}_n^{(i)}- \mathrm{p}_\mathrm{d}(L_i|\epsilon)\underline{\mathbf{W}}_n^{(i)}||_2||_p \\
	& & - ||\sum_{i=k}^{k+1}\{\mathrm{p}_\mathrm{d}(L_i|\epsilon)\overline{\mathbf{b}}_n^{(i)} - \mathrm{p}_\mathrm{d}(L_i|\epsilon)\underline{\mathbf{b}}_n^{(i)}\}||_p \IEEEyessubnumber\label{r2}\\
	& \geq & -\sum_{i=k}^{k+1}\{\mathrm{p}_\mathrm{d}(L_i|\epsilon)(||||\overline{\mathbf{W}}_n^{(i)}||_q+||\underline{\mathbf{W}}_n^{(i)}||_q||_p \\
	& & + ||||\overline{\mathbf{W}}_n^{(i)}-\underline{\mathbf{W}}_n^{(i)}||_2||_p + ||\overline{\mathbf{b}}_n^{(i)}-\underline{\mathbf{b}}_n^{(i)}||_p)\} \IEEEyessubnumber\label{r3}\\
	& = & \mathrm{p}_\mathrm{d}(L_k|\epsilon)r_k + \mathrm{p}_\mathrm{d}(L_{k+1}|\epsilon)r_{k+1}. \IEEEyessubnumber
\end{IEEEeqnarray}}}

\bibliographystyle{IEEEtran}
\bibliography{IEEEabrv,Reference}

\end{document}